\documentclass[12pt]{article}
\usepackage[margin=1in]{geometry} 
\usepackage{setspace}
\usepackage{placeins}
\usepackage{etoolbox}
\usepackage{amsmath}
\usepackage{amssymb}
\usepackage{tikz}
\usetikzlibrary{calc}
\usepackage{graphicx}
\usepackage{booktabs}  % For professional-looking tables
\usepackage{longtable} % For tables spanning multiple pages
\usepackage{tabularx}  % For adjustable column widths
\usepackage{natbib}
\usepackage{color}
\usepackage{ulem}
\usepackage{appendix}
\usepackage{enumitem}
\usepackage{comment}
\usepackage{url}
\newtheorem{theorem}{Theorem} 
\allowdisplaybreaks

\def\bSig\mathbf{\Sigma}

\def\ta{\tilde{a}}
\def\tb{\tilde{b}}

\def\b0{{\bf 0}}

\title{Time-Dependent Pseudo $\boldsymbol{R^2}$ for Assessing Predictive Performance in Competing Risks Data}

\author{Zian Zhuang$^{1}$,
Wen Su$^{2}$, %\email{w.su@cityu.edu.hk},
Eric Kawaguchi$^{3}$, and
Gang Li$^{1}$\\
$^{1}$Department of Biostatistics, University of California at Los Angeles \\
$^{2}$Department of Biostatistics, City University of Hong Kong, Hong Kong \\
$^{3}$Division of Biostatistics, University of Southern California \\
}

\begin{document}
\FloatBarrier        % 1. emit all in-text floats now

\clearpage 
%  This will produce the submission and review information that appears
%  right after the reference section.  Of course, it will be unknown when
%  you submit your paper, so you can either leave this out or put in
%  sample dates (these will have no effect on the fate of your paper in the
%  review process!)

%  These options will count the number of pages and provide volume
%  and date information in the upper left hand corner of the top of the
%  first page as in published papers.  The \pagerange command will only
%  work if you place the command \label{firstpage} near the beginning
%  of the document and \label{lastpage} at the end of the document, as we
%  have done in this template.

%  Again, putting a volume number and date is for your own amusement and
%  has no bearing on what actually happens to your paper!

%  This label and the label ``lastpage'' are used by the \pagerange
%  command above to give the page range for the article.  You may have
%  to process the document twice to get this to match up with what you
%  expect.  When using the referee option, this will not count the pages
%  with tables and figures.
\maketitle
\label{firstpage}

%  put the summary for your paper here

\begin{abstract}
Evaluating and validating the performance of prediction models is a fundamental task in statistics, machine learning, 
and their diverse applications. 
However, developing robust performance metrics for competing risks time-to-event data poses unique challenges. We first highlight how certain conventional predictive performance metrics, such as the C-index, Brier score, and time-dependent AUC, can yield undesirable results when comparing predictive performance between different prediction models. To address this research gap, we introduce a novel time-dependent pseudo $R^2$ measure to evaluate the predictive performance of a predictive cumulative incidence function over a restricted time domain under right-censored competing risks time-to-event data. Specifically, we first propose a population-level time-dependent pseudo $R^2$ measures for the competing risk event of interest and then define their corresponding sample versions based on right-censored competing risks time-to-event data. We investigate the asymptotic properties of the proposed measure and demonstrate its advantages over conventional metrics through comprehensive simulation studies and real data applications.

\end{abstract}

%  Please place your key words in alphabetical order, separated
%  by semicolons, with the first letter of the first word capitalized,
%  and a period at the end of the list.
%

\textbf{keywords}: Brier Score; C-index; Competing risks; Explained variance; Prediction performance; Survival models, Time-dependent AUC.

%  As usual, the \maketitle command creates the title and author/affiliations
%  display

%  If you are using the referee option, a new page, numbered page 1, will
%  start after the summary and keywords.  The page numbers thus count the
%  number of pages of your manuscript in the preferred submission style.
%  Remember, ``Normally, regular papers exceeding 25 pages and Reader Reaction
%  papers exceeding 12 pages in (the preferred style) will be returned to
%  the authors without review. The page limit includes acknowledgements,
%  references, and appendices, but not tables and figures. The page count does
%  not include the title page and abstract. A maximum of six (6) tables or
%  figures combined is often required.''

%  You may now place the substance of your manuscript here.  Please use
%  the \section, \subsection, etc commands as described in the user guide.
%  Please use \label and \ref commands to cross-reference sections, equations,
%  tables, figures, etc.
%
%  Please DO NOT attempt to reformat the style of equation numbering!
%  For that matter, please do not attempt to redefine anything!

\section{Introduction}

This paper addresses the problem of evaluating the predictive performance of prognostic models under right-censored competing risks time-to-event data, which plays a vital role in clinical decision-making and cost-effectiveness analyses.
Competing risks time-to-event data are ubiquitous in biomedical research and many other fields. There is a rich body of literature on statistical modeling of competing risks data and their applications, see, for example, \citet{putter2007tutorial, monterrubio2024review} for excellent surveys and further references of commonly used statistical models and more recent machine learning methods for competing risks data. Yet, methods for evaluating the predictive performance of prognostic models under competing risks are relatively limited, and each comes with its own set of limitations. 

%{\color{blue} (Jenna: After further reflection, I think it would be the easiest to use my previous version, even though there is some repetition in the following pararaph.) Prof. Li: The revisions have been reversed. - Jenna} 
\begin{comment}
\textbf{Some commonly used performance metrics includes discrimination measures such as concordance index (C-index) \citep{harrell1982evaluating}, time-dependent ROC curves \citep{heagerty2000time}, and D measure \citep{royston2004new}.   \citet{gail2005criteria} provided an overview of evaluation criteria for competing risks models, 
%including calibration, discrimination, accuracy, and explained variation, 
but do not address estimation under censoring. 
\citet{wolbers2009prognostic} have proposed adapted versions of C-index and D measure of prognostic separation with competing risks data. %initially introduced by \citet{royston2004new}. 
\end{comment}
\citet{gail2005criteria} provided an overview of evaluation criteria for competing risks models, but do not address estimation under censoring. \citet{wolbers2009prognostic} have proposed adapted versions of concordance index (C-index) and D measure of prognostic separation with competing risks data, initially introduced by \citet{royston2004new}. \citet{saha2010time} have extended time-dependent ROC curves to accommodate competing risks. It is important to note that the C-index 
and time-dependent ROC curves primarily assess discrimination, which are invariant to monotone transformations of predictions and therefore cannot differentiate between well-calibrated and poorly calibrated models. \citet{schoop2011quantifying} proposed an accuracy measure based on the Brier score, which accounts for calibration. Yet, as pointed out by \citet{gail2005criteria}, the Brier score aggregates the intrinsic variance of the outcome and the prediction error, and does not reduce to zero even when the prognostic model perfectly predicts the absolute risk. As a result, these metrics can sometimes yield misleading results when comparing different prognostic models, as  illustrated later in Sections 3 and 4. 

In contrast to the limited set of predictive performance metrics for competing risks data, a broader range of such metrics has been developed for right-censored data without competing risks.
These include discrimination measures such as the C-index \citep{harrell1982evaluating, uno2011c}, time-dependent ROC curves \citep{heagerty2005survival, uno2007evaluating}, and positive predictive functions \citep{moskowitz2004quantifying, zheng2008time, uno2007evaluating, chen2012predictive}; calibration measures such as the Brier score \citep{graf1999assessment, gerds2006consistent},  pseudo $R^2$ measures 
\citep{korn1990measures, schemper2000predictive,  stare2011measure, li2019prediction}, and other loss functions \citep{ royston2004new, o2005explained}.
In particular, \citet{li2019prediction} proposed a pair of complementary $R^2$-type metrics, $R^2$ and $L^2$, to evaluate the performance of a predicted survival function with right-censored time-to-event data in the absence of competing risks. Their method is straightforward to interpret, model-free, and has been shown to outperform several commonly used metrics in distinguishing among prognostic models across various settings.

\begin{figure}[p]
\centering
\includegraphics[width=1\linewidth]{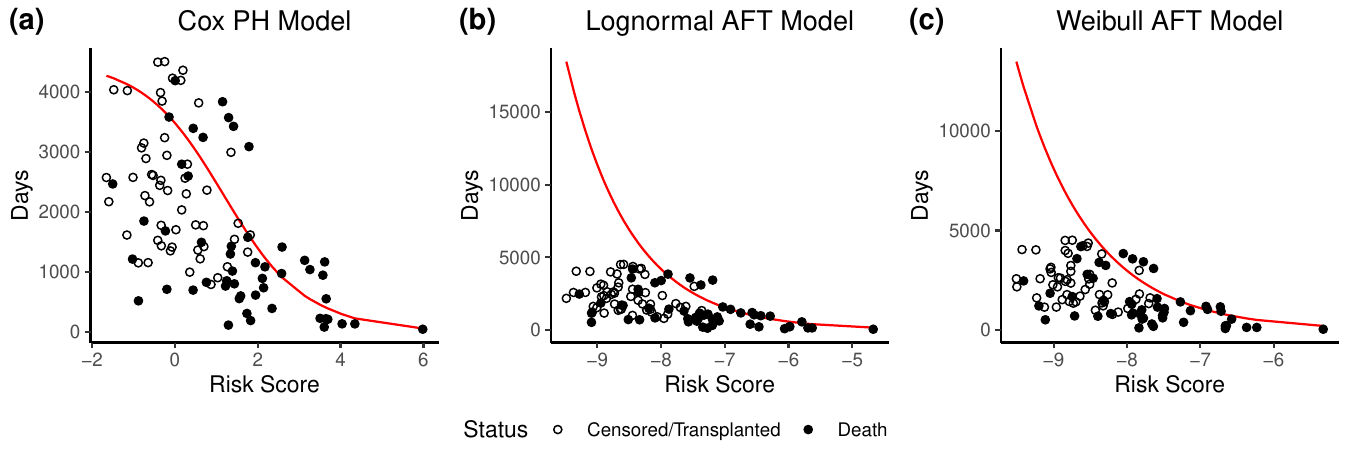}
\caption{Predicted (red solid line) and observed (dot plot) overall survival (OS) times versus the linear risk score for three prognostic models, Cox proportional hazards, Weibull AFT, and log-normal AFT, based on the Mayo Clinic primary biliary cirrhosis (PBC) dataset. The data is randomly split into training and test sets in a 2:1 ratio; the training set is used to fit the models, and the test set is used for evaluation and plotting. %Except for the Brier score, larger values of the metrics indicate better predictive performance.
}
\label{fig:pbc-intro}
\end{figure}

For instance, Figure~\ref{fig:pbc-intro} displays predicted (solid line) and observed (dot plot) overall survival (OS) times versus the linear risk score for three prognostic models, Cox's proportional hazards model, Weibull accelerated failure time (AFT) model, and log-normal AFT model, based on the Mayo Clinic primary biliary cirrhosis (PBC) dataset \citep{dickson1989prognosis}. The data is randomly split into training and test sets in a 2:1 ratio. Models are fitted on the training set and evaluated using the test set. The plots clearly show that the Cox model yields the best predictive performance.

However, as shown in Table~\ref{tab:1}, the Li--Wang method \citep{li2019prediction} is the only one that clearly identifies the Cox model as the best-performing among the three. All other metrics fail to distinguish between the models, yielding nearly identical results across all three.

\AtEndDocument{
\begin{table} 
\small
    \centering
    \caption{Predictive performance metrics across three survival models, Cox PH, Weibull AFT, and log-normal AFT, on the Mayo PBC dataset. Higher values indicate better performance, except for the Brier score where lower is better. The restricted time horizon $\tau$ is set as the maximum observed time for metrics that require it. AUC and Brier score are reported as the averages taken over 10 evenly spaced quantiles of the observed event times.}
    %\label{tab:model_comparison}
    \label{tab:1}
    \begin{tabular}{l|ccc}
        \hline\hline
        \textbf{Predictive Performance Metric} & \textbf{Cox PH} & \textbf{Weibull AFT} & \textbf{Log-normal AFT}  \\
        \hline
        Pseudo $R^2 = R^2 \times L^2$ \citep{li2019prediction} & \textbf{0.30} & \textbf{0.04} & \textbf{0.02}  \\ \hline
        Harrell's C \citep{harrell1982evaluating} & 0.81 & 0.82 & 0.82  \\ \hline
        Uno's C \citep{uno2011c} & 0.79 & 0.79 & 0.79  \\ \hline
           Brier score \citep{graf1999assessment} & 0.13  & 0.12  & 0.12  \\ \hline
        Time-dependent AUC \citep{heagerty2000time} & 0.87  & 0.88  & 0.87  \\ \hline
      %  Brier score \citep{graf1999assessment} & 0.13 (Avg. 0.12) & 0.13 (Avg. 0.11) & 0.12 (Avg. 0.11) \\ \hline
       % Time-dependent AUC \citep{heagerty2000time} & 0.89 (Avg. 0.88) & 0.90 (Avg. 0.89) & 0.90 (Avg. 0.88)  \\ \hline
        $R^2_{\mathrm{SPH}}$ \citep{stare2011measure} & 0.60 & 0.60 & 0.59 \\ \hline
        $R^2_{\mathrm{SH}}$ \citep{schemper2000predictive} & 0.41 & NA & NA  \\ 
        \hline\hline
    \end{tabular}
\end{table}
}

Due to its appealing properties, this paper extends the work of \citet{li2019prediction} to settings involving competing risks. As noted in Section~\ref{sec:2.1}, a direct extension is not feasible because the cumulative incidence function (CIF) for the event of interest is not a proper distribution function. To address this issue, rather than evaluating the performance of a predictive CIF over the entire time domain,  we shift the
focus to assessing its performance over a restricted time domain---either before a specified time horizon or at a specific time point. This approach not only enables an extension of the Li--Wang pseudo $R^2$ method to the competing risks setting, but is also practically more relevant, as event information is often unavailable beyond certain time point. Focusing on a restricted time horizon can help yield a more stable and reliable measure, especially under high-censoring, as illustrated and discussed at the end of Section \ref{sec:sample}. The resulting pseudo $R^2$ is novel even in settings with complete data, without competing risks or censoring.

Our main contributions are summarized as follows:
\begin{itemize}[leftmargin=*, labelsep=0.5em]
\item[1.] We first introduce a novel time-dependent pseudo $R^2$ metric to evaluate the performance of a predictive CIF before a specified time horizon $\tau$, by applying the Li--Wang pseudo $R^2$ framework to a working restricted event time in the competing risks setting. Specifically, we first propose a population-level time-dependent pseudo $R^2$ metric for the competing risk event of interest, and then derive a corresponding sample version based on right-censored competing risks time-to-event data.  At the population level, empirical investigations show that the proposed pseudo $R^2$ exhibits more reliable operating characteristics compared to the existing metrics such as C-index, Brier score and time-dependent AUC. For the sample version, we establish its consistency and asymptotic normality, and assess its finite-sample performance across a range of scenarios through simulation studies. Lastly, it is worth noting that the proposed time-dependent pseudo $R^2$ is novel even in settings without competing risks and with complete data. It also helps address several shortcomings associated with the original pseudo $R^2$. For example, unlike the original $R^2$ definition, the time-dependent pseudo $R^2$ is insensitive to outliers. Moreover, restricting the evaluation to an earlier time horizon can help to improve the estimation error of the sample pseudo $R^2$ in scenarios with high censoring, as illustrated in Section 3 (Figure \ref{fig:5}).

\item[2.] Analogously, we derive a novel pseudo $R^2$ metric—available in both population and sample versions—to evaluate the performance of a predictive CIF at a specified time horizon $\tau$, by applying the Li--Wang pseudo $R^2$ framework to a working event time indicator at time $\tau$ in the competing risks setting. The resulting pseudo $R^2$ provides an appealing alternative to existing time-dependent metrics, such as the Brier score and time-dependent ROC curves, for assessing the accuracy and discriminative power of a competing risks model at a fixed time $\tau$. Again, at the population level, our empirical studies demonstrate that the proposed pseudo $R^2$ possesses more desirable and robust operating characteristics than the Brier score and time-dependent AUC. For the sample version, we establish its consistency and asymptotic normality, and investigate its finite-sample estimation performance under a variety of settings through simulations.

\item[3.] We demonstrate through simulations and real-world data examples that several commonly used performance metrics for competing risks data possess inherent limitations and may produce misleading assessments under specific scenarios—phenomena that, to our knowledge, have not been previously documented in the literature. In contrast, the proposed time-dependent pseudo $R^2$ consistently exhibits robust and interpretable operating characteristics across a range of settings.
%, with empirical results conforming to theoretical and intuitive expectations.

\end{itemize}

The rest of this paper is organized as follows. In Section 2, we first review the Li--Wang pseudo $R^2$ approach for right-censored data without competing risks, and then derive novel time-dependent pseudo $R^2$ metrics for competing risks outcomes over a restricted time domain—either before a specified time horizon or at a specific time point. We derive the pseudo $R^2$ metrics both at the population level and for right-censored competing risks data, and establish the consistency and asymptotic normality of the sample version. Section 3 presents simulation studies that evaluate the operating characteristics of the proposed method in comparison to some common existing predictive evaluation metrics for competing risks data. In Section 4, we illustrate the proposed approach on two real-world medical datasets: the Mayo Clinic primary biliary cholangitis trial and the United Network of Organ Sharing data. Concluding remarks are provided in Section 5. Additional simulation results and an extended case study on the Framingham Heart dataset are provided in the Supplementary Materials. %three real-world medical datasets: the Mayo Clinic primary biliary cholangitis trial, the Framingham Heart Study, and the United Network of Organ Sharing data. Concluding remarks are provided in Section 5. Proofs of the theorems and additional simulation results are included in the Supplementary Materials.

\section{Methodology}
\label{s:model}
\subsection{Preliminaries}
\label{sec:2.1}
For reader's convenience, we first provide a brief review of the Li--Wang pseudo $R^2$ \citep{li2019prediction} for settings without competing risks.

\subsubsection{Population pseudo $R^2$} \label{sec:population}
Let $Y$ denote the outcome variable and $X$ be the associated $p$-dimensional covariate vector for a randomly selected subject from the test population. 
Let $F(y) = P(Y \leq y)$ denote the unknown marginal distribution function of $Y$.  
Given $X$, let $F^{*}(y | X)$ denote a known predictive distribution function for $F(y)$, typically obtained from a separately trained model. Let $F(y | x)=P(Y\le y | X=x)$ denote the unknown true conditional distribution function of $Y$ given $X=x$.

To evaluate how accurately $F^{*}(\cdot | X)$ predicts $F(\cdot)$, let
\begin{eqnarray}
\mu^*(X) &=& E(Y|X; F^*)= \int y \, dF^{*}(y | X), \label{eqn: mu*}\\
\mu^*_c(X) &=& \ta + \tb\mu^*(X),
\quad\mbox{$(\tilde{a}, \tilde{b})=\arg \min_{\alpha, \beta} E\{ Y -  (\alpha +\beta \mu^*(X))\}^2$}\nonumber,
\end{eqnarray}
denote the predicted value and the linearly corrected predicted value for $Y$ based on $F^{*}(\cdot | X)$, respectively.

\citet[equations (4) and (5)]{li2019prediction} established the following variance and prediction error decompositions: 
\begin{eqnarray}
\label{decomposition1} 
{var}(Y) = E\left\{\mu^*_c(X) - E(Y)\right\}^2 + E\left\{Y - \mu^*_c(X)\right\}^2,
%& & \mbox{explained variance } & \mbox{unexplained variance}
\end{eqnarray}
and
\begin{eqnarray}
\label{decomposition2}
E\left\{Y - \mu^*(X)\right\}^2 = E\left\{Y - \mu^*_c(X)\right\}^2 + E\left\{\mu^*_c(X) - \mu^*(X)\right\}^2.
\end{eqnarray}

The decompositions \eqref{decomposition1} and \eqref{decomposition2} motivate two complementary summary measures:
\begin{eqnarray}
\label{rsquare}
\rho^2 &=& \frac{E\left\{\mu^*_c(X) - E(Y)\right\}^2}{{var}(Y)}, \\
\label{lsquare}
\lambda^2 &=& \frac{E\left\{Y\!\! -\! \mu^*_c(X)\right\}^2}{E\left\{Y\!\! -\! \mu^*(X)\right\}^2},
\end{eqnarray}
where $\rho^2$, the proportion of variance in $Y$ explained by $\mu^*_c(X)$,  measures how accurately $\mu^*_c(X)$ predicts $Y$, whereas $\lambda^2$, the proportion of prediction error of $\mu^*(X)$ explained by $\mu^*_c(X)$, quantifies the discrepancy between $\mu^*(X)$ and $\mu^*_c(X)$. 
When used jointly, they provide a comprehensive evaluation of how well $\mu^*(X)$ predicts $Y$, and consequently, how accurately $F^{*}(\cdot | X)$ approximates $F(\cdot)$.

From now on, we will refer to
\begin{eqnarray*}
%\label{pseudoR}
\rho_{\text{pseudo}}^2 = \rho^2 \times \lambda^2
\end{eqnarray*}
as the population pseudo $R^2$, quantifying how accurately $F^{*}(\cdot | X)$ approximates $F(\cdot)$.

It can be shown that \citep[Theorem 2.1(c)]{li2019prediction} if the predictive distribution function is correctly specified, i.e., 
$F^{*}(y | x) = F( y | x)$ for all $x$ and $y$, then $\lambda^2= 1$, $\rho^2 =\rho^2_{NP}$, and consequently,
\begin{equation*}
%\label{nonparametricR}
\rho_{\text{pseudo}}^2 =\rho^2_{NP},
\end{equation*}
where $\rho^2_{NP} \equiv \frac{\operatorname{var}(E(Y|X))}{\operatorname{var}(Y)}$ is the nonparametric coefficient of determination, representing the proportion of variance in $Y$ explained by $E(Y | X)$. Therefore, $\rho_{\text{pseudo}}^2$ generalizes $\rho^2_{NP}$ by extending its applicability from correctly specified predictive distribution settings to scenarios where the predictive distribution may be misspecified.

\subsubsection{Sample pseudo $R^2$ with right-censored survival data without competing risks}
\citet[Section 3]{li2019prediction} also proposed sample versions of $\rho^2$ and $\lambda^2$ for right-censored survival data without competing risks based on weighted sample variance and prediction decompositions, and established their consistency and asymptotic normality.

\subsection{Time-dependent pseudo $R^2$ for competing risks data}
%cumulative incidence up to a specified time horizon}
Now consider a competing risks outcome $(Y, D)$, where $Y$ represents the time to event, and $D=k$ indicates that an event of type $k$ has occurred, with $1\le k\le K$ and $K\ge 2$. Let $X$ be a $p$-dimensional covariate vector. Without loss of generality, assume that $D = 1$ represents the event of interest. 

Let $F_1(y) = P(Y \leq y, D=1)$ denote the marginal cumulative incidence function (CIF) of type 1 event by time $y$.
Let $F_1^{*}(y | X)$ denote a predictive CIF for $F_1(y)$, typically obtained from a separately trained model. The objective is to assess how accurately $F_1^{*}(\cdot | X)$
predicts $F_1(\cdot)$.

Because the CIF's $F_1(\cdot)$ and $F_1^{*}(\cdot | X)$ are not proper distribution functions, directly extending Li--Wang pseudo $R^2$ method to the competing risks setting is challenging.  For example, there is no direct analog to the variance decomposition in \eqref{decomposition1}. To address this issue, we shift the focus from evaluating the predictive CIF $F^*_1(y | X)$ over the full time domain to assessing its predictive accuracy within a restricted time domain---either before or exactly at a specified time horizon, as discussed in the following sections.

\subsubsection{Time-dependent pseudo $R^2$ for cumulative incidence before a specified time horizon}\label{sec:intervalR2}
Let $\tau$ be a fixed time horizon.  To evaluate the predictive accuracy of $F^*_1(y | X)$ for $F_1(y)$ over the interval $[0, \tau)$,  we first derive a time-dependent pseudo $R^2$ for the competing risks outcome $(Y,D)$ at the population level.

{\sl Population time-dependent pseudo $R^2$.}  Consider  the following working $\tau$-restricted type 1 event time:
\begin{equation} \label{eq:t1tau}
Y^{(1,\tau)} = \begin{cases} Y, & \text{if } Y\le \tau \text{ and } D=1; \\
\tau, & \text{if either } \{Y>\tau\} \text{ or } \{Y\le \tau \text{ and } D\ne 1\}.
\end{cases}
\end{equation}
It can be shown that the distribution function of $Y^{(1,\tau)}$ is equal to
\begin{align*}
F^{(1,\tau)} (t)  \equiv P(Y^{(1,\tau )} \le y ) 
= \begin{cases} F_1 (y),  & \text{if } 0\le y<\tau, \\
1,  & \text{if } y\ge\tau.
\end{cases}
\end{align*}

Define the corresponding predictive distribution function as
\begin{align*}
F^{(1,\tau)*} (y|x)   
= \begin{cases} F_1^* (y|x),  & \text{if } 0\le y<\tau, \\
1,  & \text{if } y\ge\tau.
\end{cases}
\end{align*}
Then, the predictive accuracy of $F^*_1(y | X)$ for $F_1(y)$ over the interval $[0, \tau)$ directly corresponds to the predictive accuracy of $F^{(1,\tau)*}(t | X)$ for $F^{(1,\tau)}(y)$ over the entire time domain. Furthermore, since both $F^{(1,\tau)*}(y | X)$ and $F^{(1,\tau)}(y)$ are proper distribution functions, we can quantify this predictive accuracy by applying the Li--Wang pseudo $R^2$ method, as described earlier in Section \ref{sec:population}.

Specifically, we define the following population-level pseudo $R^2$ metric to evaluate the predictive accuracy of $F^*_1(y | X)$ for $F_1(y)$ over $[0, \tau)$:
\begin{eqnarray*}
%\label{restricted-pseudoR}
\rho_{\text{pseudo},1}^2([0,\tau)) = \rho_1^2([0,\tau)) \times \lambda_1^2([0,\tau)),
\end{eqnarray*}
where $\rho_1^2([0,\tau)) $ and  $\lambda_1^2([0,\tau))$ are defined by \eqref{rsquare} and \eqref{lsquare}, respectively, 
by replacing $F^*(\cdot | X)$ with $F^{(1,\tau)*}(\cdot | X)$ in equation \eqref{eqn: mu*}, and by replacing $Y$  with $Y^{(1,\tau)}$.

{\sl Sample time-dependent pseudo $R^2$ with uncensored competing risks data.}
If one observes an uncensored competing risks survival dataset, consisting of $n$ i.i.d.\ replicates of $(Y,D,X)$: 
\begin{equation*}%\label{data:cr}
\{(Y_i,D_i,X_i),\ i = 1,\ldots,n\},
\end{equation*}
then $\rho_1^2([0,\tau)) $ and $ \lambda_1^2([0,\tau))$ are consistently estimated by
\begin{equation}\label{eqn:interval R}
R^2_1([0,\tau)) =  \frac{\frac{1}{n}\sum_{i=1}^n \{\hat{\mu}_c^*(X_i)- \bar{Y}^{(1,\tau)}\}^2}{\frac{1}{n}\sum_{i=1}^n (Y^{(1,\tau)}_i -\bar{Y}^{(1,\tau)})^2},
\end{equation}
and
\begin{equation}\label{eqn:interval L}
L^2_1([0,\tau))=\frac{\frac{1}{n}\sum_{i=1}^n \{ Y^{(1,\tau)}_i -
\hat{\mu}_c^*(X_i)\}^2}{\frac{1}{n}\sum_{i=1}^n  \{Y^{(1,\tau)}_i - \mu^* (X_i) \}^2},
\end{equation}
where $\bar{Y}^{(1,\tau)}=\frac{1}{n}\sum_{i=1}^n Y^{(1,\tau)}_i$, $\hat{\mu}_c^*(x)$ is the fitted  regression function from the least squares linear regression of  $Y^{(1,\tau)}_1, \ldots, Y^{(1,\tau)}_n$ on
$\mu^* (X_1), \dots, \mu^* (X_n)$, in which $\mu^* (X_i)$ is given by replacing $F^*(\cdot | X)$ with $F^{(1,\tau)*}(\cdot | X)$ in equation \eqref{eqn: mu*}.

{\sl Sample time-dependent pseudo $R^2$ with right-censored competing risks data.}
Next, we derive a sample version of $\rho_{\text{pseudo},1}^2([0,\tau))$ with the right-censored competing risks survival data, which consist of $n$ i.i.d.\ triplets: 
\begin{equation*}%\label{data:cr}
\{(T_i,\Delta_i,X_i) \equiv (Y_i \wedge C_i, D_i \cdot \delta_i , X_i),\ i = 1,\ldots,n\}.
\end{equation*}
Here, for subject $i = 1,\ldots,n$, $\delta_i =I(Y_i \le C_i)$ and $C_i$ denotes the censoring time, which is assumed to be independent of $(Y_i, D_i, X_i)$. We construct a consistent estimate of $\rho_{\text{pseudo},1}^2([0,\tau))$,  by replacing every summation involved in \eqref{eqn:interval R} and \eqref{eqn:interval L} with a weighted summation over the uncensored subjects using the inverse probability of censoring weighting (IPCW) method.  {Specifically, the weight assigned to subject $i$ is defined by:
\begin{equation}\label{weight}
w_i = \frac{\frac{\delta_i}{\hat{G} (T_i -)}}{\sum^n_{j=1} \frac{\delta_j}{\hat{G} (T_j -)}},
\end{equation}
%$$w_i = \frac{\Delta^{(1,\tau)}_i / \hat{G} (T^{(1,\tau)}_i -)}{\sum^n_{j=1} \Delta^{(1,\tau)}_j / \hat{G} (T^{(1,\tau)}_j -)},$$ 
where $\hat{G}$ is the Kaplan-Meier estimate \citep{kaplan1958nonparametric} of $G(c) = P(C>c)$.
We then define 
$$
R_{\text{pseudo},1}^2([0,\tau)) =R_1^2([0,\tau))\times L_1^2([0,\tau)),
$$
where
\begin{equation*}
R^2_1([0,\tau)) =  \frac{\sum_{i=1}^nw_i \{\hat{\mu}_{wc}^*(X_i)- \bar{Y}_w^{(1,\tau)}\}^2}{\sum_{i=1}^nw_i (Y^{(1,\tau)}_i -\bar{Y}_w^{(1,\tau)})^2},
\end{equation*}
and
\begin{equation*}
L^2_1([0,\tau))=\frac{\sum_{i=1}^nw_i \{Y^{(1,\tau)}_i -
\hat{\mu}_{wc}^*(X_i)\}^2}{\sum_{i=1}^nw_i  \{Y^{(1,\tau)}_i - \mu^* (X_i)\}^2}.
\end{equation*}
Note that the working random variable $Y^{(1,\tau)}_i$ defined by \eqref{eq:t1tau} is observed for an uncensored subject with  $\delta_i=1$.} Here $\bar{Y}_w^{(1,\tau)} = \sum^n_{i=1} w_i Y^{(1,\tau)}_i$,  $\hat{\mu}_{wc}^{*}(x)$ is the fitted regression function from the weighted least squares of $Y^{(1,\tau)}_1, \ldots, Y^{(1,\tau)}_n$ on $\mu_{1}^* (X_1), \dots, \mu^* (X_n)$ with weighting $W=\operatorname{diag}(w_1,\dots,w_n)$. 

Similar to  Theorem 3.1 of \citet{li2019prediction}, we have the following asymptotic results.
\noindent
\begin{theorem} \label{thm1}
%Assume conditions (C1)-(C4) from the Appendix \ref{sec:conditions} hold. 
Under mild regularity conditions, as $n\rightarrow \infty$,
\begin{itemize}
\item[(a)] (Consistency) 
 $R_1^2([0,\tau)) \xrightarrow{P} \rho_1^2([0,\tau)),\text{ and } L_1^2([0,\tau)) \xrightarrow{P} \lambda_1^2([0,\tau));$

\item[(b)] (Asymptotic normality)
$$\sqrt{n} \left(R_1^2([0,\tau)) - \rho_1^2([0,\tau)) \right) \xrightarrow{d} N\left(0,\nu^2_\rho ([0,\tau)) \right),$$
and
$$\sqrt{n} \left(L_1^2([0,\tau)) - \lambda_1^2([0,\tau))\right) \xrightarrow{d} N\left(0, \nu^2_\lambda ([0,\tau)) \right),$$
\indent  where $\nu^2_\rho ([0,\tau))$ and $\nu^2_\lambda ([0,\tau))$ are the asymptotic variances.
\end{itemize}
\end{theorem}
The proof of Theorem~\ref{thm1} closely parallels the proof of Theorem 3.1 in \citet{li2019prediction}, with a subtle distinction: we assume that the predictive distribution is obtained from an independent training dataset, whereas \citet{li2019prediction} derive it from the same data used for evaluation. As a result, our asymptotic analysis is conditional on a fixed predictive distribution, which simplifies the derivation. Details are therefore omitted.

\subsubsection{Time-dependent pseudo $R^2$ at a specific time point}
To evaluate the predictive accuracy of $F^*_1(y | X)$ for $F_1(y)$, at pre-specified time point $\tau$, we
define the following  working binary outcome variable:
\begin{equation*} 
\xi^{(1,\tau)} = I( Y \le \tau, D=1).
\end{equation*}
Then,
\begin{align*}
P(\xi^{(1,\tau)} = 1) = F_{1} (\tau)
\end{align*}
and 
\begin{align*}
\mu^*(X) = E(\xi^{(1,\tau)}|X; F^*_1) =F_{1}^* (\tau|X).
\end{align*}

Therefore, replacing $Y$ by $\xi^{(1,\tau)}$ in Section \ref{sec:population} leads to the following population-level pseudo $R^2$ metric for evaluating the predictive accuracy of $F^*_1(\tau | X)$ for $F_1(\tau)$:
\begin{eqnarray*}
%\label{restricted-pseudoR point}
\rho_{\text{pseudo},1}^2(\{\tau\}) = \rho_1^2(\{\tau\}) \times \lambda_1^2(\{\tau\}),
\end{eqnarray*}
where $\rho_1^2(\{\tau\}) $ and  $\lambda_1^2(\{\tau\})$ are defined in \eqref{rsquare} and \eqref{lsquare}, by replacing $Y$ and $\mu^*(X)$  with $\xi^{(1,\tau)}$ and $F_{1}^* (\tau)$, respectively.

{\sl Sample time-dependent pseudo $R^2$ with right-censored competing risks data.} %Next,   we derive  a sample version of $\rho_{\text{pseudo},1}^2(\{\tau\})$ with the right-censored competing risks survival data in (\ref{data:cr}). 
Similar to Section~\ref{sec:intervalR2}, we derive a sample version of $\rho_{\text{pseudo},1}^2(\tau)$ by first constructing it under uncensored competing risks data, and then extending it to right-censored competing risks survival data using the IPCW method. The resulting  estimator of $\rho_{\text{pseudo},1}^2(\tau)$ is defined by:
\begin{eqnarray*}
%\label{restricted-pseudoR-point-sample}
R_{\text{pseudo},1}^2(\{\tau\}) = R_1^2(\{\tau\}) \times R_1^2(\{\tau\}),
\end{eqnarray*}
where
\begin{equation*}
R^2_1(\{\tau\}) =  \frac{\sum_{i=1}^nw_i \{\hat{\mu}_{wc}^*(X_i)- \bar{\xi}_w^{(1,\tau)}\}^2}{\sum_{i=1}^nw_i (\xi^{(1,\tau)}_i -\bar{\xi}_w^{(1,\tau)})^2},
\end{equation*}
and
\begin{equation*}
L^2_1(\{\tau\})=\frac{\sum_{i=1}^nw_i \{\xi^{(1,\tau)}_i -
\hat{\mu}_{wc}^*(X_i)\}^2}{\sum_{i=1}^nw_i  \{\xi^{(1,\tau)}_i - F_{1}^* (\tau|X_i)\}^2}.
\end{equation*}
Here, $\bar{\xi}_w^{(1,\tau)} = \sum^n_{i=1} w_i \xi^{(1,\tau)}_i$, and $\hat{\mu}_{wc}^{*}(x)$ is the fitted regression function obtained via weighted least squares of $\xi^{(1,\tau)}_1, \ldots, \xi^{(1,\tau)}_n$ on $F_{1}^* (\tau|X_1), \dots, F_{1}^* (\tau|X_n)$ with weighting $W=\operatorname{diag}(w_1,\dots,w_n)$, and the weights are defined by \eqref{weight}. Note that the working random variable $\xi^{(1,\tau)}_i$ defined by \eqref{eq:t1tau} is observed for an uncensored subject with  $\delta_i=1$.

Similar to  Theorem 3.1 of \citet{li2019prediction} and Theorem~\ref{thm1}, we have the following asymptotic results.
\noindent
\begin{theorem} \label{thm2}
%Assume conditions (C1)-(C4) from the Appendix \ref{sec:conditions} hold. 
Under mild regularity conditions, as $n\rightarrow \infty$,
\begin{itemize}
\item[(a)] (Consistency) 
 $R_1^2(\{\tau\}) \xrightarrow{P} \rho_1^2(\{\tau\}),\text{ and } L_1^2(\{\tau\}) \xrightarrow{P} \lambda_1^2(\{\tau\});$

\item[(b)] (Asymptotic normality)
$$\sqrt{n} \left(R_1^2(\{\tau\}) - \rho_1^2(\{\tau\}) \right) \xrightarrow{d} N\left(0,\nu^2_\rho (\{\tau\}) \right), $$
and
$$\sqrt{n} \left(L_1^2(\{\tau\}) - \lambda_1^2(\{\tau\})\right) \xrightarrow{d} N\left(0, \nu^2_\lambda (\{\tau\}) \right),$$
\indent  where $\nu^2_\rho (\{\tau\})$ and $\nu^2_\lambda (\{\tau\})$ are the asymptotic variances.
\end{itemize}
\end{theorem}
The proof of Theorem~\ref{thm2} is omitted, as it closely parallels the proof of Theorem 3.1 in \citet{li2019prediction}.

\subsection{Software}
A user-friendly R package, \textbf{TimeMetric}, implementing the pseudo $R^2$ and other commonly used evaluation metrics discussed in this paper, is publicly available on GitHub: (\url{https://github.com/toz015/PAmeasure}).

\section{Simulation Studies}
\label{s:sim}

We have conducted simulation studies to evaluate the performance of the proposed pseudo $R^2$ measures ($\rho_{\text{pseudo},1}^2([0,\tau))$) and ($\rho_{\text{pseudo},1}^2(\{\tau\})$) under various settings. Because the findings are similar across both metrics, we focus on presenting the results for $\rho_{\text{pseudo},1}^2([0,\tau))$ in this section. Results for $\rho_{\text{pseudo},1}^2(\{\tau\})$ are provided in the Supplmentary Materials for completeness. For a comprehensive comparison, we also evaluate the performance of several widely used prediction metrics, including the Brier score \citep{wu2018quantifying}, time-dependent AUC \citep{zheng2012evaluating}, and the C-index \citep{wolbers2014concordance}. To ensure that all prediction metrics consistently assess performance over a given time interval rather than at a single time point, we compute the AUC and Brier score at 10 evenly spaced quantiles of the observed event times and report their average values. We first evaluate all considered metrics at the population level in Section \ref{sim:population}, and then investigate the finite-sample performance of the proposed pseudo $R^2$ with right-censored samples under various scenarios and present the results in Section \ref{sec:sample}.

\subsection{Data Generation}
\label{s:data}
We generate competing risks data based on a cause-specific hazards Cox model (CSH-Cox) \citep{prentice1978analysis}, where each event type follows a Weibull distribution conditional on covariates. Specifically, we define the hazard function for event type $k$ ($k = 1,2$) as: $h_k(t; X_i) = v \lambda_k^v t^{v-1} \exp(X_i^\top\beta_k),$ where $\lambda_k$ is the scale parameter, $v$ is the shape parameter common to both event types, and $\beta_k$ is the regression coefficient vector associated with covariates $X_i$. Given the cause-specific hazard function, we can have the corresponding cumulative cause-specific hazard $H_k(t; X_i) = (\lambda_k t)^v \exp(X_i^\top \beta_k)
$. The cumulative incidence function (CIF) for event type $k$ conditional on covariates $X_i$ is:
\begin{align}
    F_k(t; X_i) &= P(Y_i \leq t, D_i = k \mid X_i)\notag\\
    &= \int_0^t h_k(u; X_i) \exp\left\{-\sum_k H_k(u; X_i)\right\}\, du. \label{eqn:CIF}
\end{align}
To generate competing risk data from the above model, we use a method similar to Li--Wang pseudo $R^2$ method. For subject $i$, we take the following steps:
\begin{itemize}[leftmargin=*, labelsep=0.5em]
\item[1.] Generate the covariate vector $X_i \sim N(\mathbf{0}, \mathbf{I})$, where $\mathbf{I}$ is the $2 \times 2$ identity matrix.
\item[2.] Generate two independent event times $T_{i1}$ and $T_{i2}$ for each subject for event type 1 and 2, based on inverse transform sampling from their respective cause-specific distributions:
\begin{align*}
Y_{ik} = H_k^{-1}(u; X_i) = \lambda_k^{-1} \left[-\log(U_{ik}) \exp(-X_i^\top \beta_k)\right]^{1/v}, \quad k=1,2,
\end{align*}
where $U_{ik} \sim \text{Uniform}(0,1)$.

\item[3.] Determine the event time $Y_i$ and event type $D_i$ for each subject as follows:
\begin{align*}
Y_i = \min(Y_{i1}, Y_{i2}), \quad
D_i = \begin{cases}
1 & \text{if } Y_i = Y_{i1}, \\
2 & \text{if } Y_i = Y_{i2}.
\end{cases}
\end{align*}

\item[4.] 
%To achieve a prespecified proportion $p$ of event type 1 in the training set (with a fixed scale parameter $\lambda_1$ for event type 1), 
Fix $\lambda_1$, repeat steps 2 and 3 to solve for $\lambda_2$ such that the resulting dataset matches the desired proportion $p$ for event type 1. The estimated $\lambda_2$ is then used to generate both the training and test datasets under the same scenario.
\end{itemize}

To incorporate censoring into the data, we generate censoring times according to a pre-specified mean censoring rate $\pi_c$. For each observation $i$, we take the following steps:
\begin{itemize}[leftmargin=*, labelsep=0.5em]
\item[1.] Draw a temporary log-censoring time $r_i$ from a normal distribution centered at zero, with the same standard deviation as the log event time.
\item[2.] Using the pre-specified proportion of censored events $\pi_c$, determine a censoring shift $\mu$ by solving for the appropriate threshold.
\item[3.] Compute the final censoring time as $C_i=\exp(\mu+r_i)$. 
\item[4.] The observed event time is calculated as \( T_i = \min(Y_i, C_i) \), where $Y_i$ is the event time.
\end{itemize}

\subsection{Population level evaluation} \label{sim:population}
\subsubsection{Simulation 1: Operating characteristics of prediction accuracy metrics} %Effect of Parameters on Prediction Metrics Under the True Model}
\label{sim1}
We evaluate and compare the operating characteristics of the proposed $\rho_{\text{pseudo},1}^2([0,\tau))$  with several widely used prediction performance metrics at the population level, under a variety of scenarios, assuming that the predictive type 1 CIF is correctly specified. Since $\rho_{\text{pseudo},1}^2([0,\tau))= \rho^2_{NP}$, for the correctly specified predictive type 1 CIF, the nonparametric $R^2$ for $Y^{(1,\tau)}$ can serve as a benchmark for evaluating the performance of the metrics under consideration. Under each scenario, the predictive type 1 CIF $F_1^*(\cdot|X)$ is obtained by fitting the cause-specific hazard Cox model with a large training competing risks dataset of size $n=10000$ without censoring. The population-level $\rho_{\text{pseudo},1}^2([0,\tau))$ for the predictive CIF $F_1^*(\cdot|X)$ is then approximated as the average of $\rho_{\text{pseudo},1}^2([0,\tau))$ over 100 independent Monte Carlo test uncensored competing risks datasets of size $n=5000$. Using the true CIF in \eqref{eqn:CIF} for prediction, we calculate population-level prediction performance metrics, including $\rho_{\text{pseudo},1}^2([0,\tau))$, C-index, Brier score, and AUC.  %Since this setting involved no censoring and evaluation was conducted at the population level, we used the nonparametric \( \rho^2_{NP} \) measure as a benchmark. 
We explore a variety of scenarios by varying the following parameters: 
\begin{itemize}[leftmargin=*, labelsep=0.5em]
\item \( v = (0.5,\ 0.75,\ 1,\ 3,\ 5,\ 10) \);
\item \( \beta_1 = (\mathbf{0.5}^\top,\ \mathbf{0.75}^\top,\ \mathbf{1}^\top,\ \mathbf{1.5}^\top) \) and \( \beta_2 = -0.2\beta_1 \);
%\item \( \beta_1 = [\beta_{11}, \beta_{12}]^T \text{ and } \beta_2 = -\beta_1, \text{ where } \beta_{11}=\beta_{12}=(0.1, 0.25, 0.5, 0.75, 1, 3, 5) \); 
\item \( p = (0.01,\ 0.05,\ 0.1,\ 0.2,\ 0.3,\ 0.4,\ 0.5,\ 0.7,\ 0.9,\ 0.95,\ 0.99) \).
\end{itemize}
The Weibull shape parameter $v$ controls the variance of event times, with larger values of $v$ corresponding to smaller variances. The regression coefficients $\beta_1$ and $\beta_2$ represent the underlying effect size for event type 1 and 2, while $p$ determines the type 1 event.
To evaluate the influence of these key parameters on performance metrics, we vary one parameter at a time while holding all other parameters constant at their default values: \( \lambda_1 = 0.5 \), \( p = 0.7 \), \( \beta_1 = [1,1]^\top \), \( v = 10 \), with the restricted time \( \tau \) set to the maximum of all event times. %\textcolor{red}{The results are summarized in Table 1 of the supplementary material. Comment: Better use ref to refer to tables/figures in the paper. Currently, we don't have supplementary materials, may need to revise.} %To better visualize the behavior of these metrics, we illustrate their performance in Figure \ref{fig:1}.

\begin{figure}[p]
    \centering
    \includegraphics[width=1\linewidth]{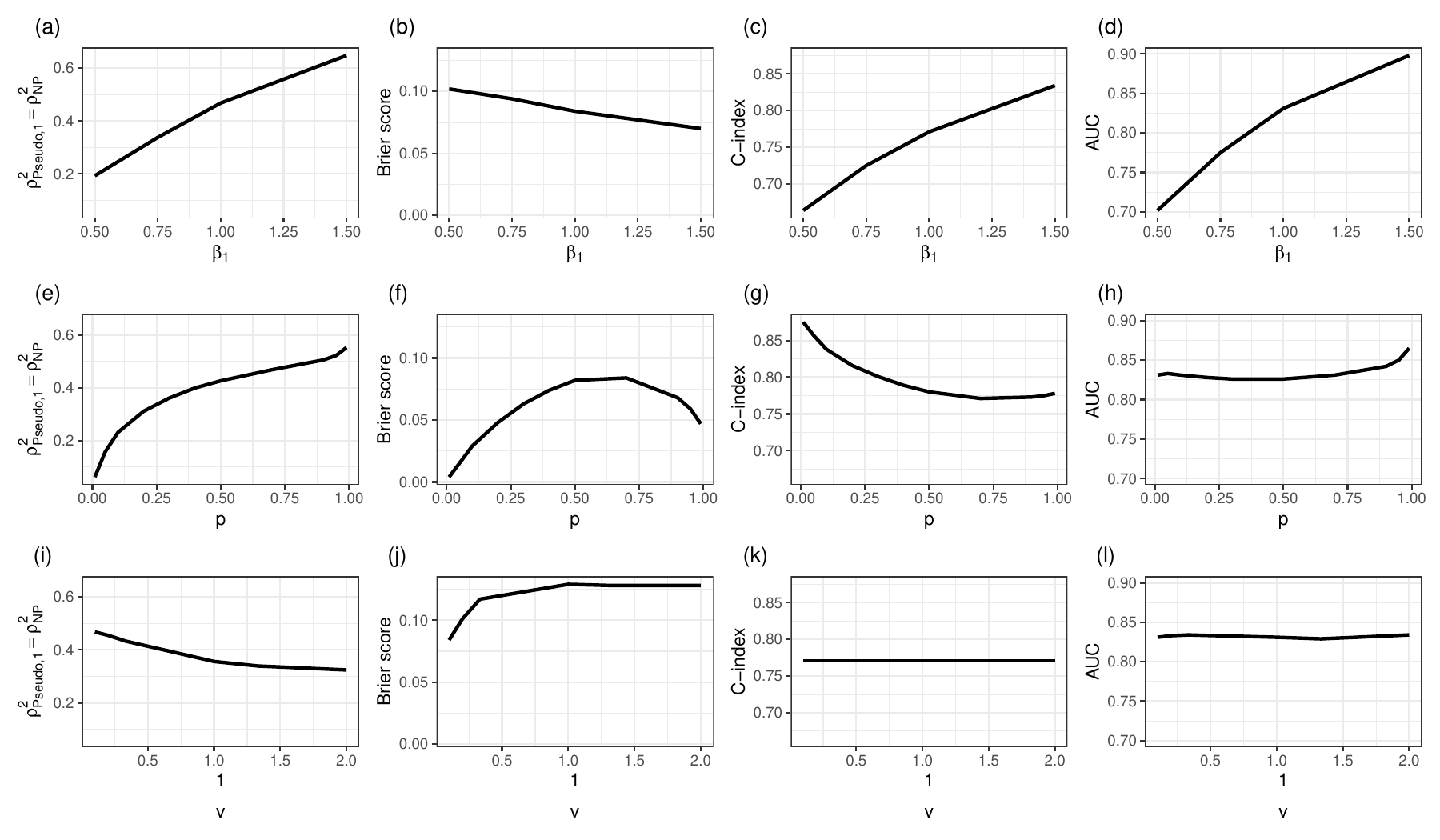}
    \caption{%Mean values in Population Simulations for Different Prediction Evaluation Methods: This figure shows the 
    Population evaluation metrics ($\rho_{\text{pseudo},1}^2([0,\tau))$, C-index, Brier score, and AUC) averaged over 100 replications in population simulations.  The data are generated from the cause-specific hazards Cox (CSH-Cox) model, and predictions are obtained using the true model. The first row shows results when varying the values of the regression coefficient $\beta_1$ for the type 1 event, while holding all other parameters fixed. The second row examines the effect of changing the proportion of event type 1 (\( p \)). The third row presents the impact of adjusting \( v \), which controls the variance of event time. When varying one parameter, all other parameters remain fixed at their default values: \( p = 0.7 \), \( \beta_1 = [1,1]^T \) and \( v = 10 \).}
    \label{fig:1}
    
\end{figure}

Figure \ref{fig:1} displays the population-level simulations results for various prediction evaluation methods. Here, $\rho_{\text{pseudo},1}^2([0,\tau))$
serves as the comparison benchmark because it equals the nonparametric \( \rho^2_{NP} \) under the correctly specified model as noted in the last paragraph of Section \ref{sec:2.1}. Except for the Brier score, higher metric values indicate better predictive performance. First, we observe that $\rho_{\text{pseudo},1}^2([0,\tau))$ measure consistently trends in the expected directions across all evaluated scenarios. For example, when the regression coefficient $\beta_1$ increases, $\rho_{\text{pseudo},1}^2([0,\tau))$ also increases; all other methods similarly increase, indicating their ability to differentiate between settings with varying $\beta_1$ values.  When the proportion of the type 1 event increases, $\rho_{\text{pseudo},1}^2([0,\tau))$ also increases. In contrast, the C-index shows an opposite trend, decreasing as the event proportion increases. Meanwhile, the Brier score peaks around an event proportion of 0.5. When event proportion exceeds 0.5, the score decreases, which aligns with the correct direction of predictive performance. However, when the event proportion is below 0.5, the Brier score increases, which goes to wrong direction. This behavior arises because the Brier score contains both the intrinsic variance of the outcome and the prediction error. In scenarios with low event proportions, the outcome variance tends to dominate, leading to misleadingly higher Brier scores  when the predictions are better. For AUC, its value decreases as the event proportion increases when the proportion of the event of interest is low. These observations suggest that changes in event proportion may reduce the ability of certain metrics to accurately reflect model prediction performance. Furthermore, when inverse of the Weibull parameter $v$ increases, (i.e., the variance of event times increases), we observe that $\rho_{\text{pseudo},1}^2([0,\tau))$ decreases and Brier score increases. C-index remains unchanged and the AUC shows only a minimal decrease, which 
indicates that these measures have limited ability to capture the model calibration with different event time distributions.

\subsubsection{Simulation 2: Comparing Prediction Performance Between Different Predictive Models}
Next, we evaluate the capability of various metrics to distinguish between different prediction methods applied to the same data.  We consider two types of cause-specific hazard (CSH) models \citep{prentice1978analysis}: the Cox proportional hazards model (CSH-Cox) and the Weibull accelerated failure–time model (CSH-AFT), together with a random survival forest \citep{ishwaran2014random} and the Fine--Gray sub-distribution hazards model \citep{fine1999proportional}.  For the CSH-AFT model, we consider a linear model for log time $Y_k$ of event type \(k\;(k = 1,2)\) with covariate \( X\in\mathbb R^{2}\),
\begin{equation*}
    \log Y_{k }= \mu_{k} +  \gamma_k^{\top}  X + \sigma_k   W,
\end{equation*}
where $\mu_{k}$ is intercept for cause $k$, $\gamma_{k}$ is the regression coefficient vector, $\sigma_{k}$ is the scale parameter associated with error $W$, which follows the standard extreme value distribution. 
%The log extreme value error implies a Weibull survival time, which has baseline Weibull parameters 
The survival times follow a Weibull distribtion with shape parameter 
\(\alpha_{k} = 1/\sigma_{k}\) and scale parameter \(\lambda_{k}= \exp(-\mu_{k}/\sigma_{k})\). %Hence the cause–specific hazard for the event type $k$ is
%\begin{equation*}
%    h_k(t; \mathbf X) = \exp(-\gamma_k^{\top}\mathbf X) %h_{0k}(t\exp(-\gamma_k^{\top}\mathbf X)),
%\end{equation*}
%where the Weibull baseline hazard, \(h_{0k}(x)\) is  $\lambda_k\alpha x^{\alpha-1}$. 
Additionally, we consider a CSH-AFT model with fixed error scale parameters \( \sigma_{1} = \sigma_{2} = \sigma = 5 \) to intentionally induce model misspecification. For each method, we also consider a reduced model with one covariate.

\begin{figure}[p]
    \centering
    \includegraphics[width=1\linewidth]{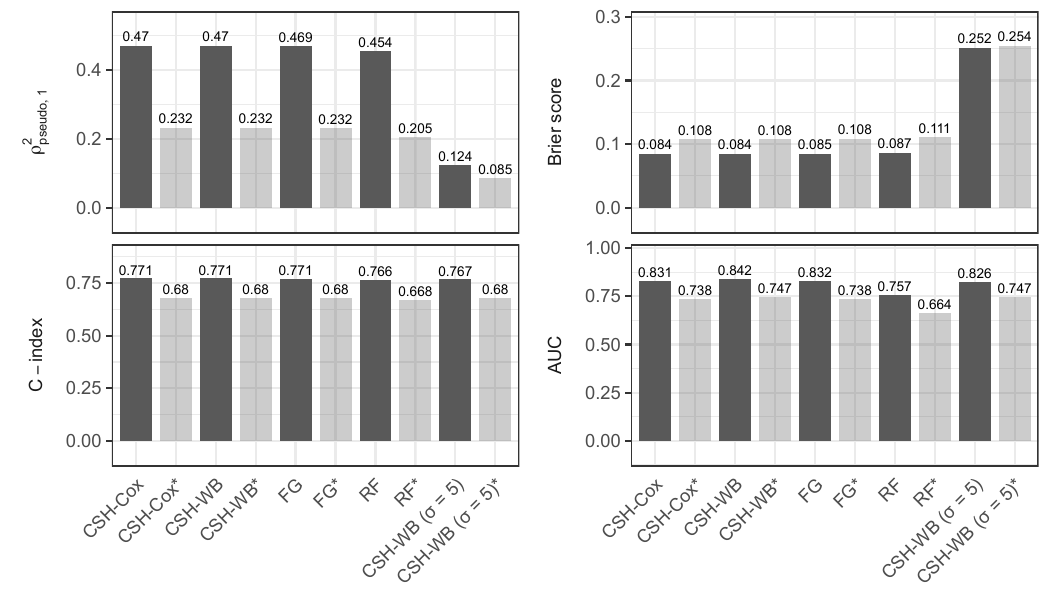}
    \caption{%Comparison of Model Prediction Performance Across Different Evaluation Methods: This plot compares the 
    Population evaluation metrics ($\rho_{\text{pseudo},1}^2([0,\tau))$, C-index, Brier score, and AUC) averaged over 100 replications for different predictive models. %Predictive performance of different models using various evaluation metrics.
    Data are generated from the cause-specific hazards Cox model, using following parameter settings: \( \lambda_1 = 0.5 \), \( p = 0.7 \), \( \beta_1 = [1,1]^\top \) and \( v = 10 \). Each subplot represents a different evaluation method, displaying results for different model types, including both the full model and the reduced model. Within every subplot the models are ordered from left to right by descending \( \rho_{\text{pseudo},1}^2([0,\tau)) \).
    %The cause-specific hazard Cox model is abbreviated as CSH-Cox, the cause-specific hazard Weibull AFT model as CSH-WB, and the cause-specific hazard Weibull AFT model with scale fixed as 5 as CSH-WB ($\sigma=5$). The Fine--Gray model is abbreviated as FG. The random-survival forest model is abbreviated as RF. Models with reduced covariates are marked with $^*$.
    The models are abbreviated as follows: CSH-Cox (cause-specific hazard Cox model), CSH-WB (cause-specific hazard Weibull AFT model), CSH-WB ($\sigma=5$) (cause-specific hazard Weibull AFT model with scale fixed at 5), FG (Fine--Gray model), and RF (random-survival forest model). Models fitted with reduced covariates are marked with $^*$.}
    \label{fig:2}
\end{figure}

We generate the data from the full cause-specific hazards Cox model, using the following parameter settings: \( \lambda_1 = 0.5 \), \( p = 0.7 \), \( \beta_1 = [1,1]^\top \), and \( v = 10 \). First, we generate a large uncensored dataset ($n=10000$) to train all models. Then, we generate 100 independent Monte Carlo samples of size 5000 as test sets, also without censoring. Prediction performance metrics are calculated for each test set based on the estimated CIF in \eqref{eqn:CIF}. The results are illustrated in Figure \ref{fig:2}. 
%Figure \ref{fig:2} compares model prediction performance across different evaluation methods. In this case, 
%The Fine--Gray model (the true model) is expected to demonstrate the best performance, and full models should outperform their nested counterparts. Our findings indicate that all evaluation metrics successfully distinguished between full and nested models. However, the C-index and AUC methods produced nearly identical prediction accuracy measures across all six models, failing to differentiate between the prediction performance of different model types.
All evaluation metrics demonstrates that full models consistently outperform their nested counterparts, in line with expectations. While the CSH-Cox (the true model) is expected to perform the best, the C-index and AUC produce contradictory results by showing similar values across all six models, failing to differentiate between the prediction performance of different model types. This behavior is expected, because the CSH-AFT model can be viewed as a monotonic transformation of a linear predictor, and both the C-index and AUC are invariant to these monotonic transformations. 

%\subsection{Finite sample performance} \label{sec:sample}
\subsection{Simulation 3: Finite sample performance of $R_{\text{pseudo},1}^2([0,\tau))$}\label{sec:sample}
\begin{figure}[p]
    \centering
    \includegraphics[width=1\linewidth]{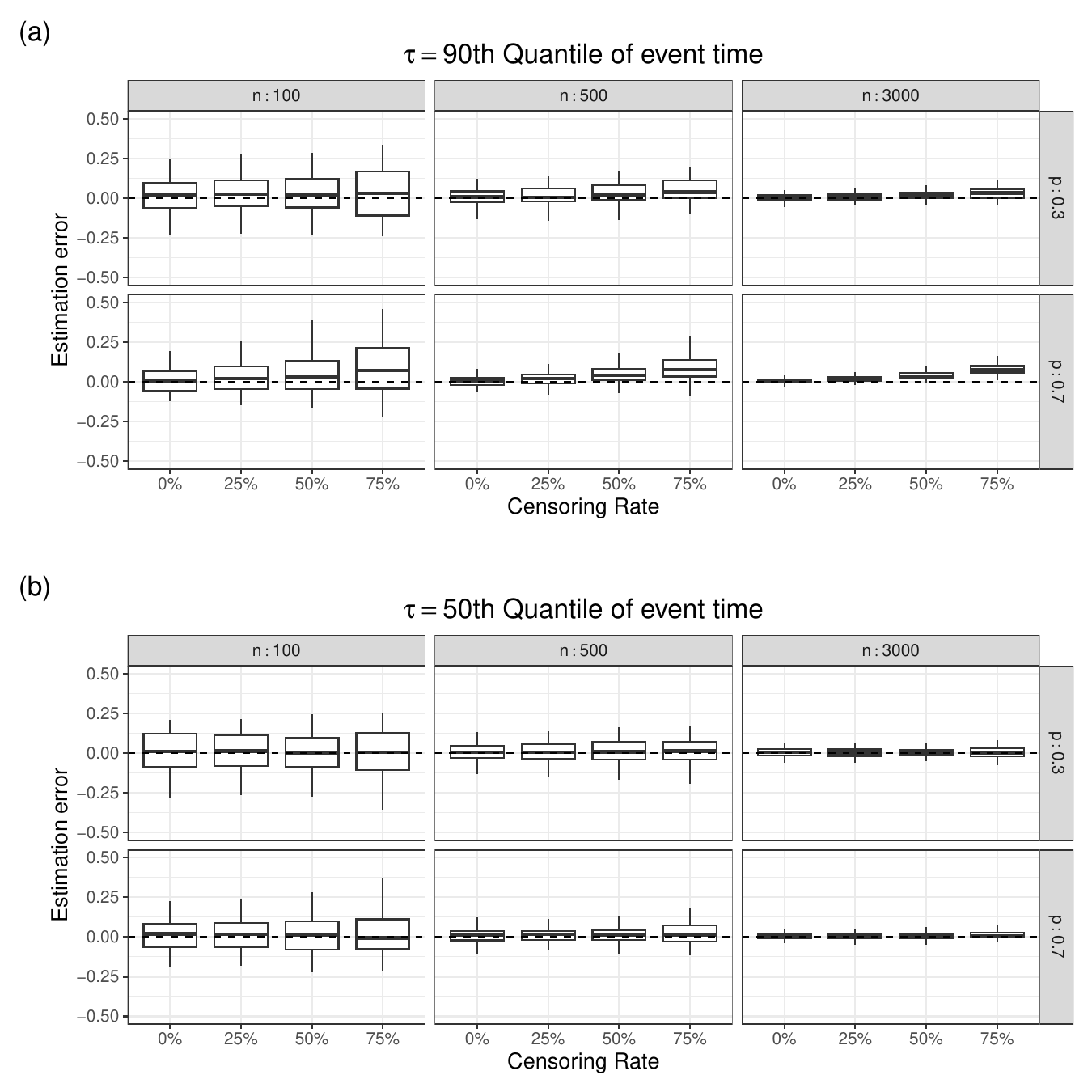}
    \caption{%Bias of Pseudo \( R^2 \) Under Different Censoring Levels: This plot illustrates the 
    Box plots of the estimation error of the sample pseudo $R^2$ ($R_{\text{pseudo},1}^2([0,\tau))$) based on 100 replicates, across different  restricted times (\( \tau = 90th \text{ and }  50th \text{ quantile of the event time}\)), different sample sizes (\( n \), each column) and different proportions of the event of interest (\( p \), each row). Within each panel, we examine the effect of changing censoring rate. Data are generated from the cause-specific hazards Cox model, using following parameter settings: \( \lambda_1 = 0.5 \), \( \beta_1 = [1,1]^\top \), and \( v = 10 \).}
    \label{fig:5}
\end{figure}

This simulation evaluates the finite-sample performance of the proposed $R_{\text{pseudo},1}^2([0,\tau))$ method under different  scenarios by varying the censoring rates $(0, 0.25, 0.5, 0.75, 0.90)$ and sample sizes \( n = (100, 500, 3000) \), event of interest proportions \( p = (0.3,  0.7) \), as well as restricted times \( \tau = (90th \text{ and }  50th \text{ quantile of the event time})\). Under each scenario, the predictive type 1 CIF $F_1^*(\cdot|X)$ is obtained by fitting the cause-specific hazard Cox model with a large training competing risks dataset of size $n=10000$ without censoring, as described in Simulation 1. We generate 100 censored test datasets as described in Section \ref{s:data}. %and use the trained model from Simulation 1 for calculating prediction accuracy metrics. 
All other parameters are fixed at their default values: \( \lambda_1 = 0.5 \), \( \beta_1 = [1,1]^\top \), and \( v = 10 \). 
%The population-level prediction measures, considered as the true benchmark values, are approximated using the trained model from Simulation 1 on a large uncensored test data of size $n = 5000$. 
For each right-censored test data set of size \( n = (100, 500, 3000) \), we calculate the estimation error of $R_{\text{pseudo},1}^2([0,\tau))$ as $R_{\text{pseudo},1}^2([0,\tau)) - \rho_{\text{pseudo},1}^2([0,\tau))$, where  $\rho_{\text{pseudo},1}^2([0,\tau))$ is calculated based on a large uncensored test data of size $n = 5000$.  
%For each replicate, we compute the bias of the estimates obtained from the censored samples. 
%To better compare across scenarios, we standardized the bias by dividing it by the estimated mean for each scenario. 
Figure \ref{fig:5} depicts the box plot of the estimation error of $R_{\text{pseudo},1}^2([0,\tau))$, across different restricted times (\( \tau\)), sample sizes (\( n \)) and proportions of the event of interest (\( p \)). 
The results indicate that both the estimation error and variance decrease as the event proportion ($p$) increases (comparing $p = 0.3$ with $p = 0.7$). The estimation error and variance of $R_{\text{pseudo},1}^2([0,\tau))$ tend to grow larger as the censoring rate increases, which is expected. In particular, in the case of high censoring, say 75\%, it is challenging to reduce the estimation bias to zero even with a large sample size ($n = 3000$). However, such estimation error can be reduced with a smaller restricted event time $\tau$ (comparing $\tau = 50$th quantile with $\tau = 90$th quantile). Specifically, for $\tau = 50$th quantile, both the bias and variance of $R_{\text{pseudo},1}^2([0,\tau))$ decrease toward zero as $n$ grows to 3000 across all scenarios, even in cases of high censoring (75\%).

\section{Applications}
\label{s:Real}
In this section, we illustrate the application of the proposed prediction accuracy measures using data from two medical studies: (1) Mayo Clinic Primary Biliary Cirrhosis (PBC), and (2) the United Network for Organ Sharing (UNOS) heart transplant registry.

\subsection{Mayo Clinic Primary Biliary Cirrhosis}

%The Mayo Clinic trial evaluates D-penicillamine treatment for primary biliary cirrhosis (PBC), conducted between 1974 and 1984. 
Between 1974 and 1984, the Mayo Clinic conducted a clinical trial investigating D-penicillamine as a therapeutic intervention for primary biliary cirrhosis (PBC). The study included 312 subjects, %. The cohort comprised 36 males (mean age: 56.20 years, SD: 11.49) and 276 females (mean age: 49.21 years, SD: 10.21), 
with a median follow-up of 5.04 years. The event specific proportions for transplants and deaths were 0.061 and 0.401, respectively. The dataset contains 17 covariates in total, from which we selected 5 that align with the components of the Mayo Risk Score (MRS)\citep{dickson1989prognosis}: patient age, serum bilirubin concentration, serum albumin concentration, standardized blood clotting time, and presence of edema following diuretic therapy. Serum bilirubin concentration, serum albumin concentration, and standardized blood clotting time are log-transformed for analysis. For estimating the CIF, we consider four approaches with death as the event of interest: cause-specific hazard Cox model (CSH-Cox), random-survival forest, the Fine--Gray model, and the CSH-AFT (Weibull) model with scale parameter $\sigma = 5$. %All models were built to predict death outcomes using the full covariate set. 
We further examine two variants of the Fine--Gray model: a reduced model using only age as a predictor for death outcomes, and a full model using all covariates to predict transplant outcomes.  We set $\tau = 3650$ days (approximately 10 years), corresponding to the 90\% quantile of the observed event times.
%, applying the Fine--Gray model to predict both mortality (using the reduced set) and transplant outcomes (using the full covariate set).

\begin{figure}[p]
    \centering
    \includegraphics[width=1\linewidth]{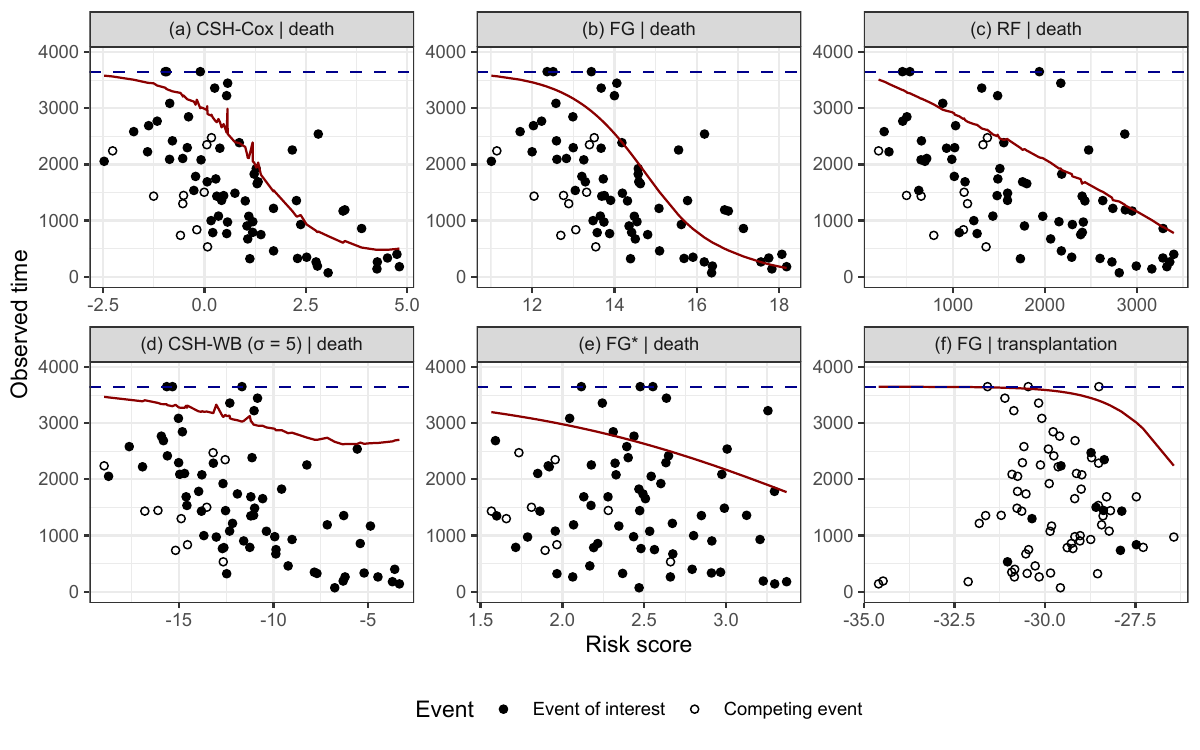}
    \caption{%PBC Data Prediction Performance Evaluation Example: The plot illustrates an example of 
    Prediction accuracy evaluation using the PBC data for several models predicting different outcomes with one sample. Here we set $\tau = 3650$ days (approximately 10 years). Death as outcome: (a) the cause-specific hazard Cox model (abbreviated as CSH-Cox), (b) the Fine--Gray model (abbreviated as FG), (c) random-survival forest model (abbreviated as RF), (d) the cause-specific hazard Weibull AFT model with scale fixed at 5 (abbreviated as CSH-WB ($\sigma=5$)), (e) the reduced Fine--Gray model (abbreviated as FG$^*$). Transplant as outcome: (f) the Fine--Gray model (abbreviated as FG). The black dots represent the event of interest, while the gray circles indicate the competing event. The red line denotes the predicted restricted mean event time for the event of interest, calculated based on the CIF.}
    \label{fig:6.2}
\end{figure}

\begin{figure}[p]
    \centering
    \includegraphics[width=1\linewidth]{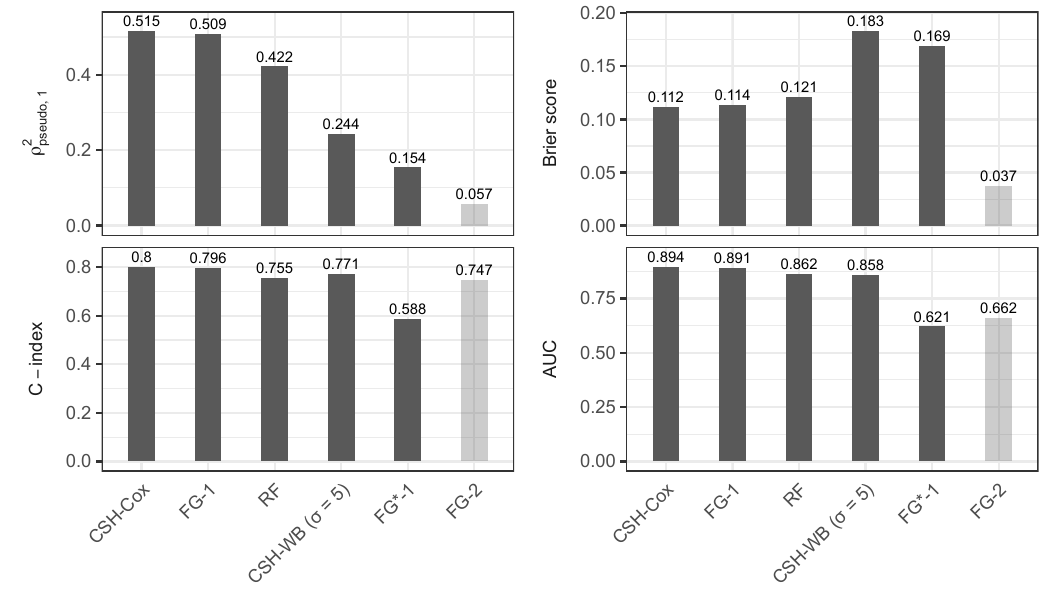}
    \caption{%Prediction Performance Evaluation Summary for the PBC Data: The plot compares several models predicting different outcomes: 
    Averaged prediction accuracy measures using the PBC data for several models predicting different outcomes over 100 replications. Each subplot shows one metric (\( \rho_{\text{pseudo},1}^2([0,\tau)) \), Brier score, C-index, or AUC) and the models within every subplot are ordered from left to right by decreasing \( \rho_{\text{pseudo},1}^2([0,\tau)) \). Here we set $\tau = 3650$ days (approximately 10 years). Death as outcome: (1) the cause-specific hazard Cox model (abbreviated as CSH-Cox), (2) the Fine--Gray model (abbreviated as FG-1), (3) random-survival forest model (abbreviated as RF), (4) the cause-specific hazard Weibull AFT model with scale fixed at 5 (abbreviated as CSH-WB ($\sigma=5$)), (5) the reduced Fine--Gray model (abbreviated as FG$^*$-1). Transplant as outcome: (6) the Fine--Gray model (abbreviated as FG-2). %The figure presents the average prediction accuracy measures, along with confidence intervals, for Pseudo \( R^2 \), Brier score, C-index, and AUC.
    }
    \label{fig:6.1}
\end{figure}

The dataset is divided into training and test sets using a 1:1 ratio. All models are trained on the training set and their predictive performance is evaluated on the test set. For evaluation, we generate 100 bootstrap samples from the test set and evaluate model performance using the pseudo $R^2$ $(\rho_{\text{pseudo},1}^2([0,\tau)))$, Brier score, C-index, and AUC. Figure \ref{fig:6.2} presents one instance of model predictions, showing subjects with observed events, both transplant and death, along with their corresponding predicted mean event times from different models. Additionally, we plot the bootstrap mean of all prediction metrics in Figure \ref{fig:6.1}.

%  random-survival forest Survival: a vector of mortality values (Ishwaran et al., 2008) representing estimated risk for each individual calibrated to the scale of the number of events

From Figure \ref{fig:6.2} and Figure \ref{fig:6.1}, we observe that for the death outcome, both the $\rho_{\text{pseudo},1}^2([0,\tau))$ and Brier score measures consistently identify the Fine--Gray  and CSH-Cox model with full covariates as having the best predictive performance, which aligns with the observed results in Figure \ref{fig:6.2}. However, while the AUC and C-index correctly distinguish between the full and reduced models, they fail to differentiate predictive performance across different model types. When comparing across outcomes, $\rho_{\text{pseudo},1}^2([0,\tau))$, AUC, and C-index all suggest that models predicting death performed better than those predicting transplant. On the contrary, the Brier score assigns a lower (better) value to the model predicting transplant outcomes, which contradicts the observed patterns in Figure \ref{fig:6.2}. Among the four metrics, only $\rho_{\text{pseudo},1}^2([0,\tau))$ yields performance assessments fully consistent with expectations.

\subsection{United Network of Organ Sharing}

In 2018, the United Network for Organ Sharing (UNOS) introduced a new heart allocation system aimed at prioritizing the sickest patients, improving waitlist outcomes, and expanding the sharing of organ donations. We consider all adult, first-time, heart-only candidates who were included on the UNOS Registry waitlist between October 18, 2018 to August 30, 2023.  A total of 16,691 individuals were included in our analysis. We classify subjects as either right-censored due loss to followup or observing one of three competing events (death/deterioration, heart transplantation, recovery) based on their removal code at time of delisting. Follow-up data was available until September 30, 2024 and candidates still on the waitlist at that time were administratively censored. The event specific proportions for death/deterioration, heart transplantation, recovery were 0.068, 0.778, and 0.032, respectively. Due to low recovery rates, we only focus on predicting cumulative incidence for death/deterioration or heart transplantation, and treat recovery events as right-censoring.

Data are split into a training ($N = 8345$) and test set ($N = 8346$). Covariates included in the models are acuity status at listing (Status 1-3 vs. Status 4-6), age at listing, biological sex, race/ethnicity, diabetes status at listing, implantable cardioverter defibrillator at listing,  history of ischemic cardiomyopathy at listing, and on dialysis at time of listing. As with previous analyses, we compare $\rho_{\text{pseudo},1}^2([0,\tau))$ with several prediction metrics for four models predicting heart transplantation: CSH-Cox model, random-survival forest, Fine--Gray model, and CSH-AFT (Weibull) model with scale parameter $\sigma = 5$. We also evaluate a reduced Fine--Gray model that excludes acuity status at listing and a Fine--Gray model that predicts death/deterioration.  %We explored two values for $\tau$, 60 and 365, corresponding to predicting early (two-month and one-year) incidence.
For all comparisons, we set $\tau = 365$ days, corresponding to the 80\% quantile of the observed event times.

\begin{figure}[p]
    \centering
    \includegraphics[width=1\linewidth]{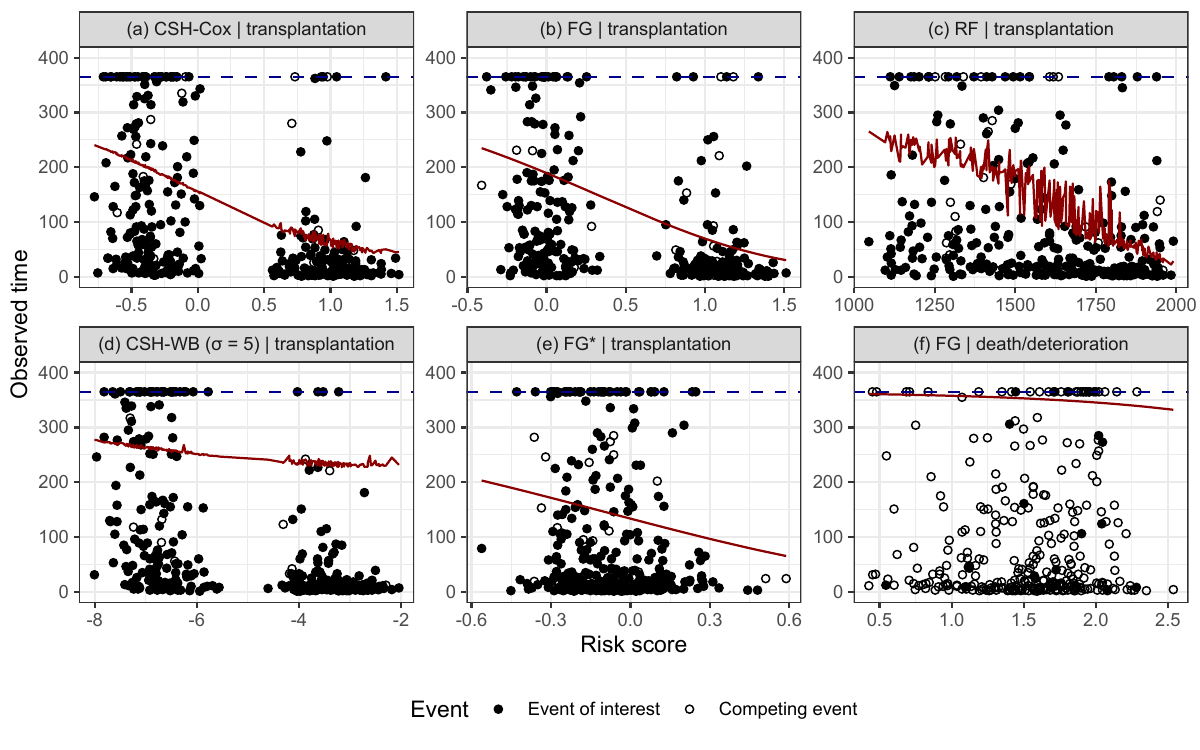}
    \caption{
    Prediction accuracy evaluation using the UNOS data for several models predicting different outcomes with one sample. Here we set $\tau = 365$ days. For illustration purpose, only 2000 observations are randomly selected from the full dataset for plotting. Heart transplantation as outcome: (a) the cause-specific hazard Cox model (abbreviated as CSH-Cox), (b) the Fine--Gray model (abbreviated as FG), (c) random-survival forest model (abbreviated as RF), (d) the cause-specific hazard Weibull AFT model with scale fixed at 5 (abbreviated as CSH-WB ($\sigma=5$)), (e) the reduced Fine--Gray model (abbreviated as FG$^*$). Death/deterioration as outcome: (f) the Fine--Gray model (abbreviated as FG). The black dots represent the event of interest, while the gray circles indicate the competing event. The red line denotes the predicted restricted mean event time for the event of interest, calculated based on the CIF.}
    \label{fig:7.2}
\end{figure}

\begin{figure}[p]
    \centering
    \includegraphics[width=1\linewidth]{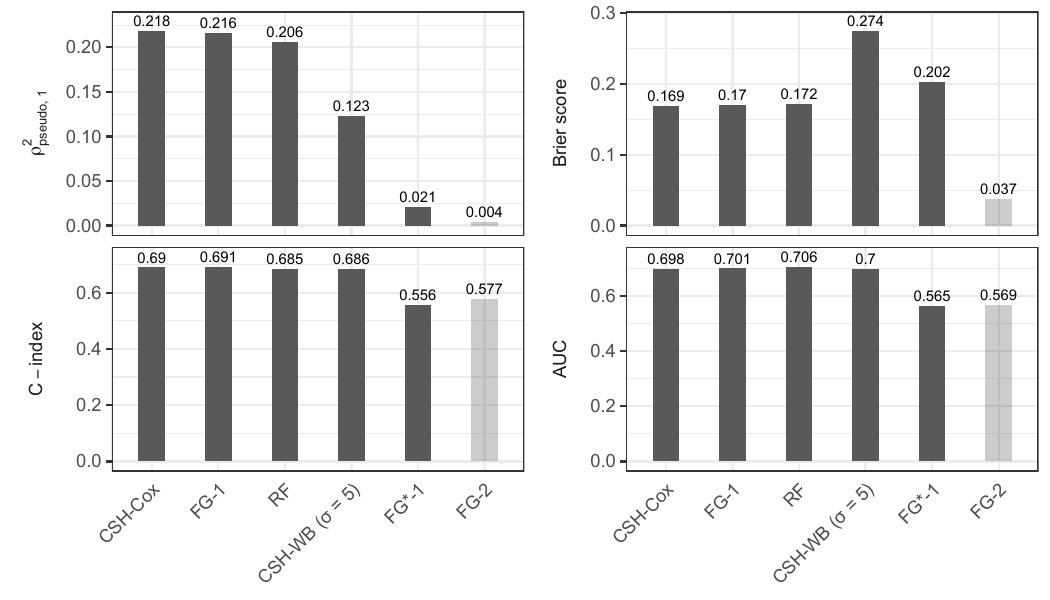}
    \caption{
    Averaged prediction accuracy measures using the UNOS data for several models predicting different outcomes over 100 replications. Each subplot shows one metric (\( \rho_{\text{pseudo},1}^2([0,\tau)) \), Brier score, C-index, or AUC) and the models within every subplot are ordered from left to right by decreasing \( \rho_{\text{pseudo},1}^2([0,\tau)) \). Here we set $\tau = 365$ days. Heart transplantation as outcome: (1) the cause-specific hazard Cox model (abbreviated as CSH-Cox), (2) the Fine--Gray model (abbreviated as FG-1), (3) random-survival forest model (abbreviated as RF), (4) the cause-specific hazard Weibull AFT model with scale fixed at 5 (abbreviated as CSH-WB ($\sigma=5$)), (5) the reduced Fine--Gray model (abbreviated as FG$^*$-1). Death/deterioration as outcome: (6) the Fine--Gray model (abbreviated as FG-2). %The figure presents the average prediction accuracy measures, along with confidence intervals, for Pseudo \( R^2 \), Brier score, C-index, and AUC.
    }
    \label{fig:7.1}
\end{figure}

Figure \ref{fig:7.2} plots the predicted survival times against the observed ones,
%illustrates an example of model predictions, 
while Figure \ref{fig:7.1} summarizes the bootstrap mean of all prediction measures across the models. First, the exclusion of acuity status at listing, a key factor of predicting outcomes, worsens predictions, and all of $\rho_{\text{pseudo},1}^2([0,\tau))$, Brier score, C-index, and AUC can distinguish the full and reduced models. Second, across all models, the $\rho_{\text{pseudo},1}^2([0,\tau))$, C-index, and AUC are higher for heart transplantation than for death/deterioration. This is expected since heart transplantation has a higher event rate than death/deterioration. However, the Brier score goes in the opposite direction, suggesting poorer calibration for heart transplantation. Lastly, when comparing across methods, CSH-Cox model, random-survival forest, and Fine--Gray model achieve the best prediction accuracy (higher $\rho_{\text{pseudo},1}^2([0,\tau))$, C-index, AUC, and lower Brier score). Nevertheless, C-index and AUC fail to flag the inferior performance of the CSH-AFT (Weibull) model, whose poor fit is apparent in Figure \ref{fig:7.2}. Among the four measures, $\rho_{\text{pseudo},1}^2([0,\tau))$ is the only one that produces performance evaluations consistent with expectations.

\section{Discussion}
\label{s:Dis}

We have developed a novel pseudo $R^2$ measure specifically designed to evaluate prediction accuracy for right-censored data in the presence of competing risks. Theoretical properties of the proposed measure have been investigated, establishing its consistency and asymptotic normality. We conducted extensive simulations to evaluate the performance of the proposed pseudo $R^2$ along with several common metrics, across a variety of scenarios. The results show that the proposed pseudo $R^2$ is the only measure that consistently exhibits robust and reliable performance across all evaluated settings. 
Notably, in cases where traditional metrics such as the C-index, Brier score, and AUC fail to distinguish the predictive performance of different predictive distributions, the pseudo $R^2$ measure successfully differentiates among them.
Furthermore, we illustrate the utility of the proposed measure and its potential advantages over competing methods using some real-world datasets including the Primary Biliary Cholangitis study and UNOS heart transplant registry. 
%While single discrimination metrics alone may not always reflect model prediction performance accurately, it is important to incorporate calibration methods in addition to discrimination measures for a more comprehensive evaluation.
%These applications highlight the practical utility and versatility of our method in diverse clinical and epidemiological settings, demonstrating its potential to enhance predictive modeling in the presence of competing risks.
%Our work provides researchers and practitioners with a rigorous and interpretable tool for assessing predictive performance for survival data, particularly in settings involving competing risks. 

For future research, it would be useful to explore extensions of the proposed pseudo $R^2$ measure to accommodate more complex survival data structures, such as interval censoring and truncation. Additionally, it would be of interest to extend the proposed pseudo $R^2$ measure to some common epidemiological designs such as nested case-control and case-cohort designs.  

\section*{Supplementary Materials}

The supplementary file contains detailed simulation results for the time-dependent pseudo $R^2$ at a specific time point, as well as an additional real data example based on the Framingham Heart dataset.

%\section*{Acknowledgments}

%\section*{Funding}
%This work is supported in part by the Research Grant Council of Hong Kong (15303319).

%\section*{Conflict of Interest}
%None declared.

%\section*{Data Availability}

%  Here, we create the bibliographic entries manually, following the
%  journal style.  If you use this method or use natbib, PLEASE PAY
%  CAREFUL ATTENTION TO THE BIBLIOGRAPHIC STYLE IN A RECENT ISSUE OF
%  THE JOURNAL AND FOLLOW IT!  Failure to follow stylistic conventions
%  just lengthens the time spend copyediting your paper and hence its
%  position in the publication queue should it be accepted.

%  We greatly prefer that you incorporate the references for your
%  article into the body of the article as we have done here
%  (you can use natbib or not as you choose) than use BiBTeX,
%  so that your article is self-contained in one file.
%  If you do use BiBTeX, please use the .bst file that comes with
%  the distribution.  In this case, replace the thebibliography
%  environment below by
%
%  \bibliographystyle{biom}
% \bibliography{mybibilo.bib}

\newpage
\bibliographystyle{plainnat} 
\bibliography{ref}  % Your .bib file without the .bib extension

%  If your paper refers to supporting web material, then you MUST
%  include this section!!  See Instructions for Authors at the journal
%  website http://www.biometrics.tibs.org

\label{lastpage}

\end{document}

\maketitle

\section{Simulation Results for $\rho_{\text{pseudo},1}^2(\{\tau\})$}\label{sup:A}

In this section, we conduct simulation studies to evaluate the performance of specific time point pseudo $R^2$ measure ($\rho_{\text{pseudo},1}^2(\{\tau\})$), as well as several widely used prediction metrics at a fixed time point, including the Brier score \citep{wu2018quantifying} and time-dependent AUC \citep{zheng2012evaluating}. Using the same data generation settings and parameter configurations as in Section 3 of the main text, we first evaluate the metrics at the population level, then examine the finite-sample performance of the proposed pseudo $R^2$ under various right-censoring scenarios.

\subsection{Population level evaluation}
\subsubsection{Simulation 1: Operating characteristics of prediction accuracy metrics.}

\begin{figure}[H]
    \centering
    \includegraphics[width=1\linewidth]{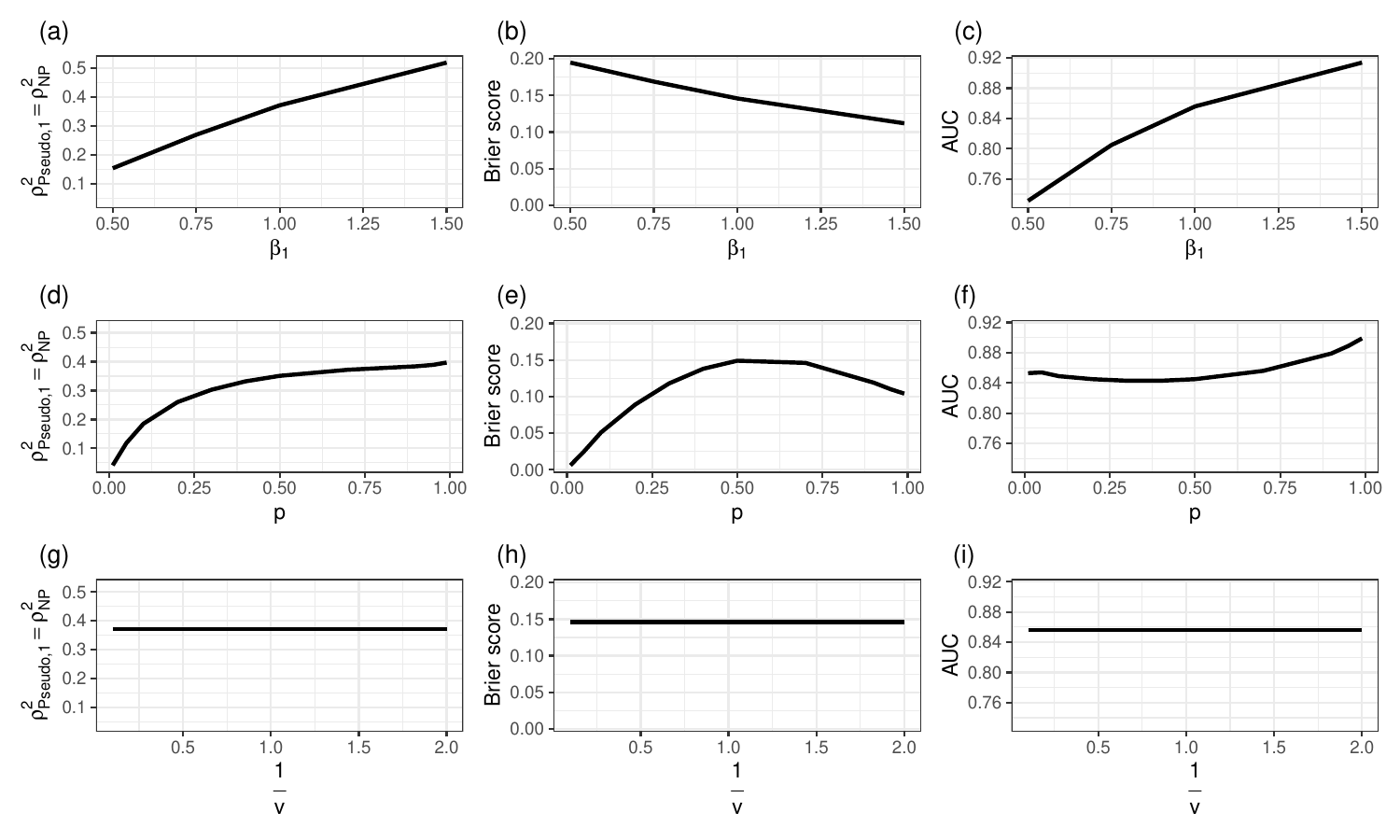}
    \caption{Population evaluation metrics ($\rho_{\text{pseudo},1}^2(\{\tau\})$, Brier score, and AUC) averaged over 100 replications in population simulations. The data are generated from the cause-specific hazards Cox (CSH-Cox) model, and predictions are obtained using the true model. The first row presents results under varying values of the regression coefficient $\beta_1$ for the type 1 event, with all other parameters fixed. The second row examines the effect of changing the proportion of event type 1 (\( p \)). The third row presents the impact of adjusting the inverse of \( v \), which controls the variance of event time. When varying one parameter, all other parameters remain fixed at their default values: \( p = 0.7 \), \( \beta_1 = [1,1]^T \) and \( v = 10 \).}
    \label{fig:pop-point1}
\end{figure}

Web Figure \ref{fig:pop-point1} presents the results of population-level simulations for various prediction evaluation metrics. The nonparametric \( \rho^2_{NP} \) measure is used as a benchmark for comparison. Except for the Brier score, higher values of each metric indicate better predictive performance. We observe that $\rho_{\text{pseudo},1}^2(\{\tau\})$ measure shows consistent and expected trends across all evaluated scenarios. For example, as the regression coefficient $\beta_1$ increases, $\rho_{\text{pseudo},1}^2(\{\tau\})$ also increases. Other metrics follow similar trends, indicating their ability to differentiate between settings with varying $\beta_1$ values.  When the proportion of type 1 events increases, $\rho_{\text{pseudo},1}^2(\{\tau\})$ also increases. However, the Brier score peaks around a 0.5 event proportion because of its dependence on both the intrinsic variance of the outcome and the prediction error. At low event proportions, the outcome variance dominates, potentially leading to inflated Brier scores even when prediction accuracy improves. As for the AUC, when proportion of the event of interest is low, its value decreases as the event proportion increases. These observations indicate that changes in event proportion may reduce the ability of certain metrics to reflect model prediction performance accurately. Lastly, when inverse of the Weibull parameter $v$ increases, (i.e., greater variability in event times), we observe that all three metrics remains unchanged. This is expected, as these metrics assess predictive performance at a fixed time point and are therefore invariant to changes in the overall distribution of event times.

\subsubsection{Simulation 2: Comparing Prediction Performance Between Different Predictive Models}

\begin{figure}[H]
    \centering
    \includegraphics[width=1\linewidth]{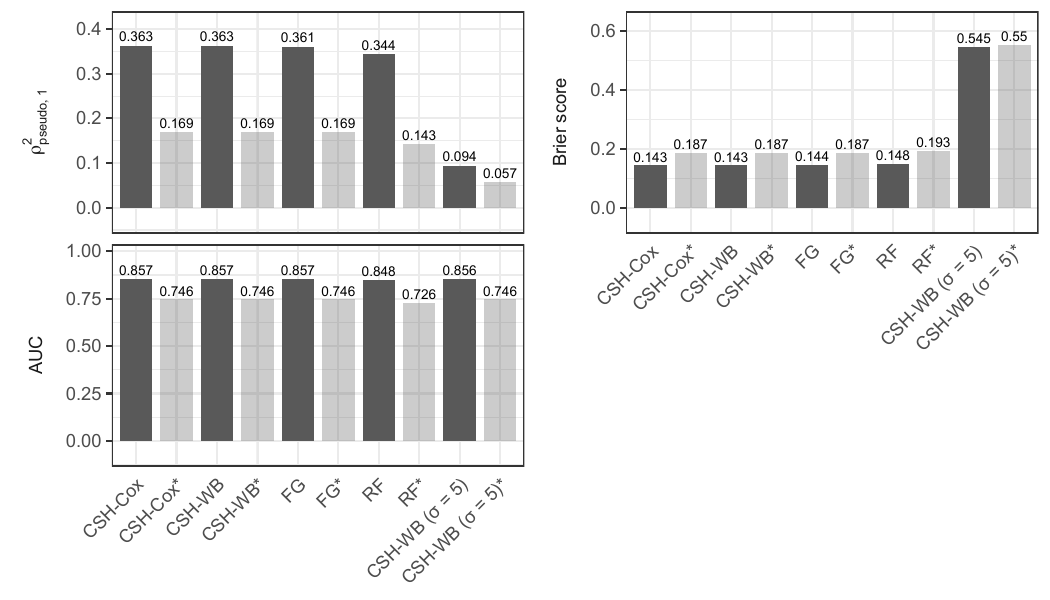}
   \caption{Population evaluation metrics ($\rho_{\text{pseudo},1}^2(\{\tau\})$, Brier score, and AUC) averaged over 100 replications for different predictive models.
    Data are generated using the following parameter settings: \( \lambda_1 = 0.5 \), \( p = 0.7 \), \( \beta_1 = [1,1]^\top \) and \( v = 10 \). Each subplot corresponds to a different evaluation method and presents results across various model types, including both the full model and the reduced model. Within every subplot the models are ordered from left to right by descending $\rho_{\text{pseudo},1}^2(\{\tau\})$. The models are abbreviated as follows: CSH-Cox (cause-specific hazard Cox model), CSH-WB (cause-specific hazard Weibull AFT model), CSH-WB ($\sigma=5$) (cause-specific hazard Weibull AFT model with scale fixed at 5), FG (Fine--Gray model), and RF (random-survival forest model). Models fitted with reduced covariates are marked with $^*$.}
    \label{fig:pop-point2}
\end{figure}

Next, we evaluate the ability of various metrics to distinguish between different prediction models applied to the same dataset. As shown in Web Figure \ref{fig:pop-point2}, all evaluation metrics consistently indicate that full models outperform their nested counterparts, aligning with expectations. While the CSH-Cox (the true model) is expected to perform the best, the AUC metric produces contradictory results. The AUC metric yields nearly identical values across all five models, failing to differentiate their predictive performance. This outcome is expected, as AUC is invariant to monotonic transformations, and the CSH-AFT model can be viewed as a monotonic transformation of the linear predictor.

\subsection{Simulation 3: Finite sample performance of $R_{\text{pseudo},1}^2(\{\tau\})$}\label{sec:sample}

\begin{figure}[H]
    \centering
    \includegraphics[width=1\linewidth]{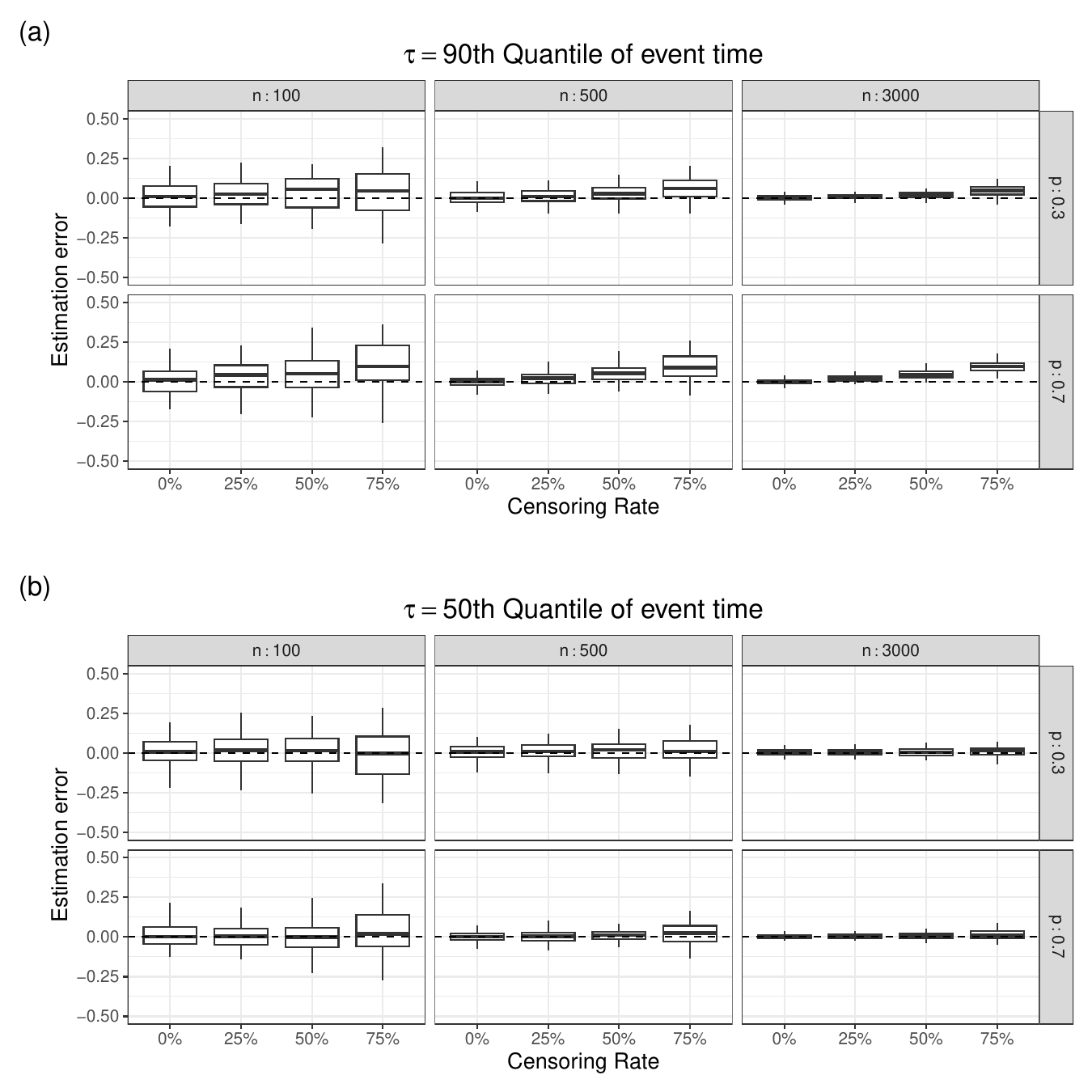}
    \caption{Box plots of the estimation error of the sample pseudo $R^2$ ($R_{\text{pseudo},1}^2(\{\tau\})$) based on 100 replicates, across different  restricted times (\( \tau = 90th \text{ and }  50th \text{ quantile of the event time}\)), different sample sizes (\( n \), each column) and different proportions of the event of interest (\( p \), each row). Within each panel, we assess the impact of varying levels of censoring. Data are generated using following parameter settings: \( \lambda_1 = 0.5 \), \( \beta_1 = [1,1]^\top \), and \( v = 10 \).}
    \label{fig:sample-point1}
\end{figure}

This simulation evaluates the finite-sample performance of the proposed $R_{\text{pseudo},1}^2(\{\tau\})$ method under different right-censoring scenarios. Web Figure \ref{fig:sample-point1} presents the box plot of the estimation error of $R_{\text{pseudo},1}^2(\{\tau\})$ across different restricted times (\(\tau\)), sample sizes (\( n \)) and proportions of the event of interest (\( p \)). The results suggest that both the estimation error and variance of the $R_{\text{pseudo},1}^2(\{\tau\})$ decrease as the event proportion (\( p \)) increases (comparing $p=0.3$ with $p=0.7$). The estimation error and variance tend to be larger for higher censoring rate, which is expected. In particular, in the case of high censoring, say 75\%, it is challenging to reduce the estimation bias to zero even with a large sample size ($n = 3000$). However, choosing a smaller restricted event time $\tau$ helps reduced this estimation error (comparing $\tau=50th$ quantile with $\tau=90th$ quantile). Notably, for $\tau=50th$ quantile, both the bias and variance of the $R_{\text{pseudo},1}^2(\{\tau\})$ approach toward zero as $n$ increases to 3000 across all scenarios, even under high censoring (75\%).

\newpage

\section{Framingham Heart Study Data Analysis}
The Framingham Heart Study is a longitudinal prospective study investigating the etiology of cardiovascular disease (CVD) and identifying risk factors 
%and their joint effects 
among residents of Framingham, Massachusetts. For our analysis, we utilize a dataset from the study (No. N01-HC-25195), which %provided with permission from the National Heart, Lung, and Blood Institute (NHLBI). 
includes 4434 participants who were examined during three separate periods occurring approximately 6 years apart from 1956 to 1968, with laboratory measurements and clinical data collected at each visit. Participants were subsequently monitored for 24 years to track health outcomes including hypertension, cardiovascular events, and mortality. We focus on hypertension and death as the primary events of interest for their high incidence rates, and use the most recent examination records prior to each event for analysis. After excluding participants with a history of hypertension at baseline, the  final dataset consists of 2783 individuals, with a median follow-up time of 6.08 years. The event specific proportions for hypertension diagnoses and deaths is 0.597 and 0.117, respectively. %1,229 males with mean age 54.49 years (SD: 9.13) and 1,554 females with mean age 53.57 years (SD: 8.20). The cohort contained 
%794 censored cases, 1,662 hypertension diagnoses, and 327 deaths,  %(IQR 3.89-12.02).
 From 14 available covariates, we select sex, age, serum total cholesterol, systolic/diastolic blood pressure, smoking status, cigarettes/day, BMI, diabetes status, heart rate, serum glucose, and education level. We estimate the CIF using four approaches: cause-specific hazard Cox model (CSH-Cox), random-survival forest model, the Fine--Gray model, and the CSH-AFT (Weibull) model with scale parameter $\sigma = 5$. All models are implemented using the full covariate set to predict hypertension outcomes. Additionally, we consider a reduced Fine--Gray model that only include three covariates: sex, serum total cholesterol, and age for predicting hypertension.  We also apply the Fine--Gray model with the full covariate set to predict death outcomes. For all predictions, we evaluate model performance at a fixed time horizon of $\tau = 5475$ days (approximately 15 years), which corresponds to the 88.3\% quantile of the observed event times.
%The models were trained and tested following the same approach as in the PBC data analysis. 

\begin{figure}
    \centering
    \includegraphics[width=1\linewidth]{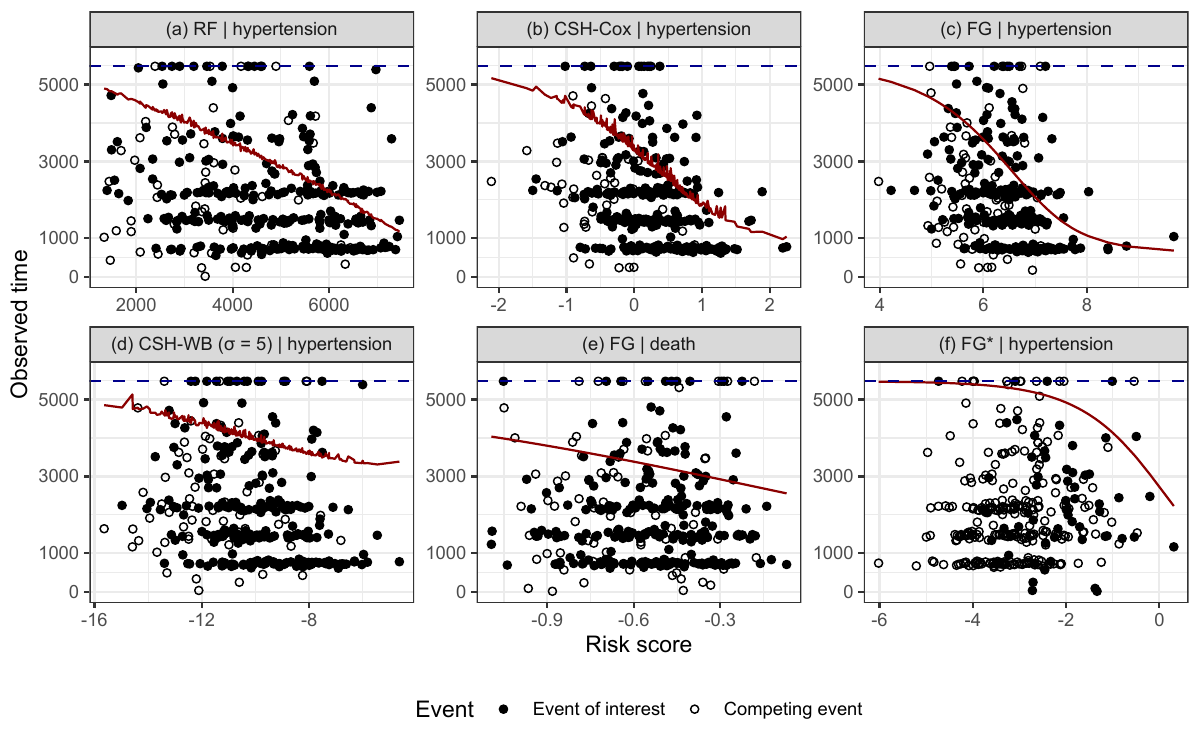}
    \caption{%Framingham Heart Study Data Prediction Performance Evaluation Example: The plot illustrates an example of evaluating prediction accuracy using Framingham Heart Study Data. It compares several models predicting different outcomes: (1) the Fine--Gray model for hypertension, (2) the random-survival forest model for hypertension, (3) the CSH-AFT (Weibull) model for hypertension, (4) the Fine--Gray model for death, and (5) the nested Fine--Gray model for hypertension. The black dots represent the event of interest, while the gray circles indicate the competing event. The red line denotes the predicted restricted mean event time for the event of interest, calculated based on the CIF.
    Prediction accuracy evaluation using the Framingham Heart Study data for several models predicting different outcomes with one sample. Here we set $\tau = 5475$ days (approximately 15 years). For illustration purpose, only 2000 observations are randomly selected from the full dataset for plotting. Hypertension as outcome:  (a) random-survival forest model (abbreviated as RF), (b) the cause-specific hazard Cox model (abbreviated as CSH-Cox), (c) the Fine--Gray model (abbreviated as FG), (d) the cause-specific hazard Weibull AFT model with scale fixed at 5 (abbreviated as CSH-WB ($\sigma=5$)), (e) the reduced Fine--Gray model (abbreviated as FG$^*$); death as outcome: (f) the Fine--Gray model (abbreviated as FG). The black dots represent the event of interest, while the gray circles indicate the competing event. The red line denotes the predicted restricted mean event time for the event of interest, calculated based on the CIF. }
    \label{fig:7.2}
\end{figure}

\begin{figure}
    \centering
    \includegraphics[width=0.9\linewidth]{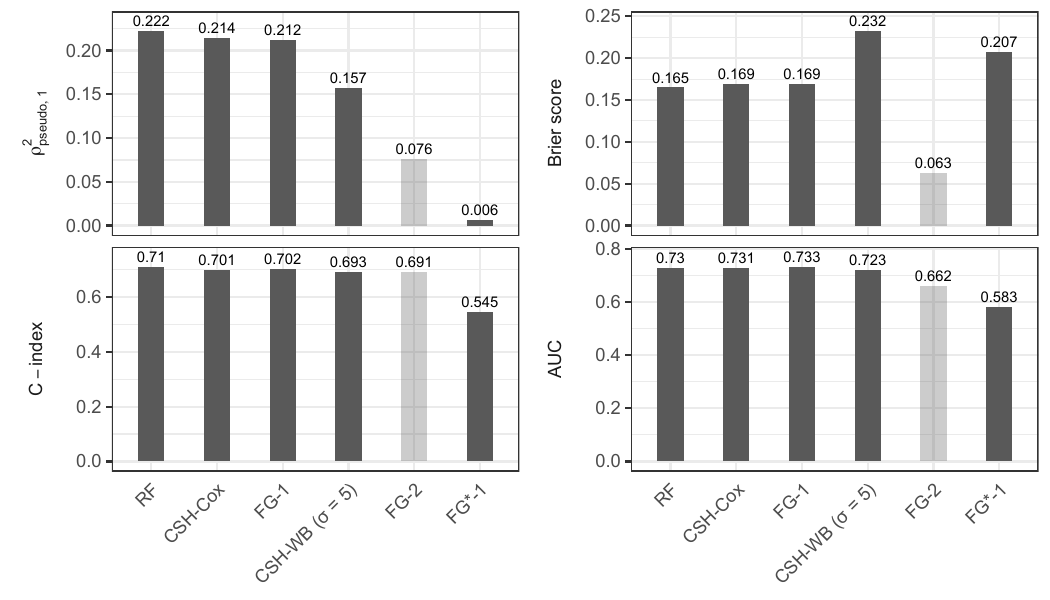}
    \caption{%Prediction Performance Evaluation Summary for the Framingham Heart Study Data: The plot compares several models predicting different outcomes: (1) the Fine--Gray model for hypertension, (2) the random-survival forest model for hypertension, (3) the CSH-AFT (Weibull) model for hypertension, (4) the Fine--Gray model for death, and (5) the nested Fine--Gray model for hypertension. The figure illustrates the average prediction accuracy measures, including confidence intervals, for Pseudo \( R^2 \), Brier score, C index and AUC.
    Averaged prediction accuracy measures using the Framingham Heart Study data for several models predicting different outcomes over 100 replications.  Each subplot shows one metric ($\rho_{\text{pseudo},1}^2(\{\tau\})$, Brier score, C-index, or AUC) and the models within every subplot are ordered from left to right by decreasing $\rho_{\text{pseudo},1}^2(\{\tau\})$. Here we set $\tau = 5475$ days (approximately 15 years). Hypertension as outcome:  (a) random-survival forest model (abbreviated as RF), (b) the cause-specific hazard Cox model (abbreviated as CSH-Cox), (c) the Fine--Gray model (abbreviated as FG-1), (d) the cause-specific hazard Weibull AFT model with scale fixed at 5 (abbreviated as CSH-WB ($\sigma=5$)), (e) the reduced Fine--Gray model (abbreviated as FG$^*$-1); death as outcome: (f) the Fine--Gray model (abbreviated as FG-2).}
    \label{fig:7.1}
\end{figure}

To evaluate predictive performance, we calculate the pseudo $R^2$ $(\rho_{\text{pseudo},1}^2([0,\tau)))$, Brier score, C-index, and AUC. Web Figure \ref{fig:7.2} illustrates an instance of model predictions, while Web Figure \ref{fig:7.1} summarizes the bootstrap mean of all prediction measures across the models. For the hypertension outcome, all measures consistently identify the random-survival forest, CSH-Cox and Fine--Gray models with full covariates as having the best predictive performance. However, both C-index and AUC indicate that the CSH-AFT (Weibull) model and the Fine--Gray model with death as outcome performed almost as well as the random-survival forest and the full Fine--Gray model, which contradicts expectations. Moreover, the Brier score suggests that the Fine--Gray model with death as outcome performs the best, which is inconsistent with the observed trends in Web Figure \ref{fig:7.2}. Of the four measures, $\rho_{\text{pseudo},1}^2([0,\tau))$ is the only metric that yielded performance evaluation results in line with intuitive expectations and consistent with the empirical patterns shown in Web Figure \ref{fig:7.2}.

%From Figures \ref{fig:7.2} and \ref{fig:7.1}, we observe that for the hypertension outcome, all four measures correctly identify the two models with the best predictive performance: the Fine--Gray model and the random-survival forest model with full covariates. However, the CSH (Weibull) model also achieves a high AUC and C-index, comparable to the Fine--Gray and random-survival forest models. The differences among these models are minor relative to the overall values of the measures. This similarity in magnitude may be misleading. Even a small difference in AUC and C-index values can correspond to significant differences in prediction accuracy. As for different outcomes, Pseudo \( R^2 \), AUC, and C-index suggest that models predicting hypertension outperform those predicting death. However, the Brier score provides a lower value to the model predicting death, which is inconsistent with the observed trends in Figure \ref{fig:7.2}.

\newpage
%\section*{References}
\bibliographystyle{apalike}
\bibliography{ref}